\documentclass[11pt,a4paper]{article}

\usepackage[margin=1in]{geometry}
\usepackage[T1]{fontenc}
\usepackage[utf8]{inputenc}
\usepackage{lmodern}
\usepackage{microtype}
\usepackage{amsmath,amssymb,amsthm}
\usepackage{booktabs}
\usepackage{tabularx}
\usepackage{multirow}
\usepackage{tikz}
\usetikzlibrary{positioning,arrows.meta,shapes.geometric,fit,backgrounds,calc,patterns}
\usepackage{pgfplots}
\pgfplotsset{compat=1.18}
\usepackage{algorithm}
\usepackage{algpseudocode}
\usepackage{enumitem}
\usepackage{xcolor}
\definecolor{accent}{HTML}{1F3A5F}
\definecolor{soft}{HTML}{E8EEF4}
\definecolor{warn}{HTML}{B8860B}
\definecolor{ok}{HTML}{2E7D32}
\definecolor{stop}{HTML}{B71C1C}
\definecolor{lavender}{HTML}{6A5ACD}
\usepackage[colorlinks=true,linkcolor=accent,urlcolor=accent,citecolor=accent]{hyperref}
\usepackage{titlesec}
\titleformat{\section}{\Large\bfseries\color{accent}}{\thesection}{1em}{}
\titleformat{\subsection}{\large\bfseries\color{accent}}{\thesubsection}{1em}{}
\titleformat{\subsubsection}{\normalsize\bfseries\color{accent}}{\thesubsubsection}{1em}{}

\newtheorem{proposition}{Proposition}
\newtheorem{definition}{Definition}
\newtheorem{example}{Example}
\newtheorem{property}{Property}
\theoremstyle{remark}
\newtheorem*{remark}{Remark}

\title{\vspace{-1.5em}\color{accent}\textbf{SARC: A Governance-by-Architecture Framework\\ for Agentic AI Systems}\\[0.3em]\large\textit{Compiling Regulatory Obligations into Runtime Constraints}}
\author{Gaston Besanson\thanks{Working paper, May 2026. Affiliation: Universidad Torcuato Di Tella. Code and reproducibility artifacts: \url{https://github.com/besanson/sarc-governance}.}}
\date{May 2026}

\begin{document}
\maketitle

\begin{abstract}
\noindent Agentic AI systems increasingly act through tools, sub-agents, and external services, but governance controls are still commonly attached to prompts, dashboards, or post-hoc documentation. This creates a structural mismatch in regulated settings: obligations that must constrain execution are often evaluated only after execution has occurred. We introduce SARC, a runtime governance architecture for tool-using agents that treats constraints as first-class specification objects alongside state, action space, and reward. A SARC specification declares each constraint's source, class, predicate, verification point, response protocol, and operating point, then compiles these declarations into four enforcement sites in the agent loop: a Pre-Action Gate, an Action-Time Monitor, a Post-Action Auditor, and an Escalation Router. We formalize the minimal invariants required for specification-trace correspondence, show why finite reward penalties do not generally substitute for hard runtime constraints, and extend the architecture to multi-agent workflows through constraint propagation, authority intersection, and attribution-preserving trace trees. We implement a prototype audit checker for finite SARC specifications and runtime traces, and report a reproducible synthetic evaluation over 50 seeds comparing SARC against post-hoc audit, output filtering, workflow rules, and policy-as-code-only baselines on a procurement task. In that task SARC executes zero hard-constraint violations under exact predicates; its declared PAA throttling response reduces soft-window overages by 89.5\% relative to the policy-as-code-only baseline, illustrating the value of explicit response placement rather than an intrinsic performance advantage of the SARC label. Predicate-noise and enforcement-failure sweeps are consistent with the claim that residual hard violations under SARC scale with enforcement-stack error rather than with environmental violation opportunity. The evaluation is intended to test architectural mechanisms under controlled assumptions, not to estimate deployment-grade procurement performance. SARC does not make agentic systems compliant or safe by itself; it provides the architectural substrate through which obligations can be made executable, inspectable, and auditable at runtime.
\end{abstract}

\noindent\textbf{Keywords:} agentic AI; AI governance; system architecture; runtime enforcement; constrained reinforcement learning; policy-as-code; EU AI Act; multi-agent systems; auditability by construction.

\section{Introduction}\label{sec:intro}

Agentic AI systems (LLM-based architectures that perceive, plan, and act through tool use over multiple steps with reduced human supervision \cite{yao2023react,schick2023toolformer,shinn2024reflexion,wang2024voyager}) are entering production in regulated environments faster than the governance infrastructure around them has matured. Two distinct conversations exist in parallel. The technical conversation is about agents: planning, tool use, memory, evaluation, scaling. The regulatory conversation is about obligations: risk management, human oversight, transparency, post-market monitoring. The two conversations rarely meet inside the same artifact. Compliance, when it happens, is produced as documentation overlaid on a technical system that was not designed to accommodate it.

\paragraph{Why now.} Recent public evaluations of frontier AI systems and LLM agents establish that the runtime location of governance controls now matters operationally. The UK AI Security Institute's 2025 Frontier AI Trends Report \cite{aisi2025trends} documents that the time horizon for autonomously completed software tasks has been doubling roughly every eight months, with frontier models in 2025 completing tasks that require human experts more than ten years of experience; METR's measurements of task-completion time horizons show consistent exponential growth on multi-hour software-engineering tasks \cite{metr2025horizon}. Frontier model system cards \cite{anthropic2025opus45card} document increasing capability for multi-step tool use, autonomous task execution, and agentic workflows in production. Cross-agent prompt-injection benchmarks \cite{debenedetti2024agentdojo} demonstrate non-trivial attack surfaces in multi-agent deployments. These developments do not by themselves imply unsafe deployment, but they make the runtime location of governance controls more important: when systems can act through tools at increasing autonomy and time horizon, governance mechanisms attached only to documentation, prompts, or post-hoc review are structurally late.

This paper shows that the two conversations should meet inside the system architecture, not inside the documentation that describes it. We introduce SARC (State, Action Space, Reward, Constraints) as a governance-by-architecture framework. The first three components are inherited directly from reinforcement learning \cite{sutton2018rl} and are used in their standard sense. The fourth, $C$, is the contribution: an explicit constraint primitive, named at the same level as $S$, $A$, and $R$, treated as a first-class architectural component, and enforced at declared points in the agent runtime.

\paragraph{Positioning.} This work is positioned as a contribution to AI systems architecture, operational governance, and regulatory engineering rather than to reinforcement learning theory. The propositions stated are scoping devices, not optimality results; the formal counterexample in \S\ref{sec:counter} addresses one place where formal precision is necessary, but the paper as a whole is a systems-and-architecture contribution to a literature \cite{schneider2003enforcement,sandhu2007rolebased,fournet2003stack,jin2017opa} that has long understood that policy enforcement belongs in the runtime.

\paragraph{What SARC contributes.} SARC connects three layers that current practice keeps separate: the regulatory obligation (a clause in a statute or contract), the system specification (the four-tuple $\langle S, A, R, C \rangle$), and the agent runtime (the loop in which the LLM perceives, plans, and acts). When $C$ is a first-class architectural component, an obligation under EU AI Act Article 14 (human oversight) compiles into a constraint declaration which compiles into a runtime check, and the audit trail is generated by the system rather than reconstructed from it. The framework sits at the intersection of three established disciplines: control systems \cite{astrom2008feedback}, where invariants are maintained as runtime properties of the controller-plant loop rather than as documentation; distributed systems, where cross-process attribution and trust boundaries are first-class concerns; and regulatory governance, where the obligation is to make conformity evidence available ex ante rather than reconstructing it only ex post.

\paragraph{Five specific contributions.} The four-tuple notation is a memorable framing, but it is not the paper's primary contribution. The contribution is a \emph{specification-to-runtime compilation discipline for regulated tool-using agents}, with five specific components.

\begin{enumerate}[leftmargin=2em,topsep=0.3em,itemsep=0.3em]
\item \emph{A constraint specification model for tool-using agents.} Each constraint is represented as a $\langle\texttt{src}, \texttt{class}, \texttt{pred}, \texttt{verif}, \texttt{resp}\rangle$ object with a declared operating point. The model is implementation-substrate-agnostic; what is fixed is the metadata required for a constraint to be executable and auditable.

\item \emph{An agent-loop enforcement architecture.} Constraints compile into four named enforcement points (PAG, ATM, PAA, ER) according to class-specific placement rules: hard constraints are enforced before or during dispatch, soft constraints over partial or completed actions, escalation constraints with a bounded human-response path relative to the action's reversibility window. The placement discipline, rather than the tuple notation, is what converts governance from documentation into execution control.

\item \emph{A checkable auditability property and a prototype checker.} \emph{Specification-trace correspondence} requires that for every action recorded in the trace, an auditor can determine which constraints were applicable, where they were evaluated, what outcome they produced, and what response was taken. We state this as a formally checkable property in \S\ref{sec:audit-checker} and implement a prototype checker that performs the four required passes over a finite specification and trace.

\item \emph{A reproducible synthetic evaluation.} We compare SARC against four baselines (post-hoc audit, output filtering, workflow rules, policy-as-code-only) on a synthetic procurement task across 50 seeds with reported 95\% confidence intervals, and run a predicate-noise / enforcement-failure sweep that supports the residual-violation scaling argument under controlled assumptions. Results are reported in \S\ref{sec:sim} and \S\ref{sec:residual}.

\item \emph{A multi-agent extension.} Constraint inheritance, authority intersection, and attribution-preserving trace trees are designed to prevent constraint laundering, authority escalation, and attribution dilution across agent boundaries.
\end{enumerate}

The differentiator from prior work is therefore measurable: \emph{coverage of action-dispatch paths, auditability of constraint evaluations, and class-typed response placement}, demonstrated on a reproducible benchmark, rather than the rhetorical move of naming a fourth primitive.

\paragraph{Outline.} \S\ref{sec:scope} positions SARC against existing decision-theoretic formalisms. \S\ref{sec:def} defines the four components, states two scoping propositions with a formal counterexample (\S\ref{sec:counter}), and specifies the eight invariants of a minimal SARC-conformant system (\S\ref{sec:invariants}). \S\ref{sec:arch} presents a reference runtime architecture with four enforcement points. \S\ref{sec:ops} characterizes operational trade-offs (latency, calibration, queueing). \S\ref{sec:compile} describes the compilation path with an Annex III mapping. \S\ref{sec:loop} presents the SARC-governed agent loop. \S\ref{sec:practitioner-compare} compares SARC to the agent-governance approaches practitioners deploy today. \S\ref{sec:multi} treats multi-agent systems substantively (topologies, recursive constraint propagation, multi-principal authority composition, the orchestrator loop, and cross-agent failure modes). \S\ref{sec:sim} reports a simulation-based illustration with sensitivity analysis. \S\ref{sec:bolted} catalogues failure modes. \S\ref{sec:economic} discusses economic implications, framing constraint calibration as a capital-allocation decision. \S\ref{sec:concl} concludes with the thesis line and an adoption ladder.

\section{SARC and Reinforcement Learning: A Note on Scope}\label{sec:scope}

Because three of SARC's four components are RL primitives, the framework is exposed to a reasonable misreading: that we are proposing yet another decision-theoretic formalism. We dispatch this misreading at the outset. SARC operates upstream of optimization machinery, not alongside it.

The classical Markov decision process \cite{puterman1994mdp} is a tuple $(S, A, P, R, \gamma)$. The constrained MDP \cite{altman1999cmdp} extends this object with a vector of cost functions and admissibility bounds. Safe and constrained reinforcement learning \cite{garcia2015safe,achiam2017cpo,ray2019benchmarking,stooke2020responsive,yang2021wcsac} operationalize CMDPs through algorithms that respect such constraints during training, deployment, or both. These are mature literatures, and we have no quarrel with any of them.

SARC operates at a prior stage. Before any optimization is well-posed (before a CMDP can be solved, a safe-RL algorithm can be selected, or a policy can be evaluated), an architect must commit to what $S$, $A$, $R$, and $C$ are for the deployment at hand. In agentic LLM systems this commitment is non-trivial: the state is partially constructed online through retrieval and tool calls; the action space is dynamically expanded through tool registration \cite{schick2023toolformer,patil2024gorilla}; the reward is rarely scalar and frequently shifts with stakeholder negotiation; and the constraints arise from regulatory, ethical, and operational sources external to the optimization problem. SARC names the four objects whose specification is necessary before any downstream optimization machinery can be applied, and provides architectural slots for them in the runtime.

\paragraph{A control-theoretic reading.} SARC can be interpreted as a closed-loop control system in which constraints define invariant sets over system trajectories, enforced at runtime rather than verified post-hoc. The agent loop is the controller; the tool registry and external systems form the plant; the constraint set $C$ encodes the invariants the controller must maintain; and the four enforcement points are the runtime mechanisms by which trajectories that would leave the invariant region are blocked, modified, or escalated. This reading recovers the standard control-systems lineage \cite{astrom2008feedback} and makes precise what is otherwise treated colloquially in agentic-AI work: SARC is the architectural commitment that invariants are runtime properties of the controller-plant loop, not documentation properties of the deployed artifact. In this interpretation, SARC enforces invariants over system trajectories rather than constraining policies ex ante, which is the structural distinction between runtime governance and policy synthesis.

A SARC specification is \emph{compilable} into a CMDP where appropriate, with the safe-RL machinery as the right tool for solving the resulting problem. SARC and CMDP are upstream and downstream of each other; they are not substitutes. The framework also has direct kinship with the policy-as-code literature \cite{jin2017opa,baier2024policymachine} and with runtime enforcement of security policies \cite{schneider2003enforcement,fournet2003stack}, whose intellectual lineage runs through the reference monitor concept \cite{anderson1972security,lampson1971protection}: a small, tamperproof, always-invoked component that mediates access to a protected resource. SARC's enforcement points are reference monitors specialized to the agent loop. The runtime-verification literature \cite{leucker2009runtime,falcone2018taxonomy,bartocci2018runtime} (particularly the cyber-physical-systems strand, where invariants must be enforced under hard real-time constraints with bounded latency) provides the closest technical analogue: SARC's enforcement points correspond to monitor placements in the runtime-verification taxonomy (PAG as inline pre-condition monitor, ATM as inline progress monitor, PAA as post-condition monitor), and the response semantics in $C$ map to enforcement primitives (suppress, modify, escalate). The framework also draws on the distributed-systems tradition for trace structure: the attribution-preserving trace tree of \S\ref{sec:multi} is, structurally, a workflow-centric distributed trace \cite{sigelman2010dapper,sambasivan2016diagnosing} adapted to agentic execution, and the orchestrator-worker isolation patterns it relies on follow the conventions of large-scale cluster managers \cite{burns2016borg}. SARC inherits these foundations and specializes them to the agent runtime: it specifies which constraint class belongs at which monitor placement, and how regulatory obligation compiles into predicate.

\subsection{Differentiation from Policy-as-Code}\label{sec:vs-pac}

The most likely reviewer objection is that SARC reduces to policy-as-code (PaC) for agents. We address this directly. Policy-as-code is an \emph{implementation substrate}: a way to express authorization, admission, and routing rules as machine-readable artifacts, evaluated by a generic decision engine such as Open Policy Agent \cite{jin2017opa}. SARC is an \emph{agent-specific specification discipline}: it determines (i) what classes of constraint exist for an agentic system ($C_h$, $C_s$, $C_e$), (ii) which point in the agent loop is the appropriate enforcement site for each class (PAG, ATM, PAA, ER), (iii) how regulatory obligations compile into those constraints (\S\ref{sec:compile}), and (iv) what runtime contracts the loop must honor (Algorithm~\ref{alg:loop}). PaC is a candidate substrate for implementing $C$ at the policy or tool layer; it is not a substitute for the SARC tuple, the loop placement, the reversibility-window-aware queueing model of \S\ref{sec:queue}, or the regulatory compilation path. A PaC engine can evaluate a SARC predicate; it cannot tell you which predicate to write, where in the agent runtime to evaluate it, or what the response semantics should be when it fires.

\subsection{Comparison to Path-Policy Runtime Governance}\label{sec:vs-paths}

The closest contemporary work is Kaptein, Khan, and Podstavnychy's \emph{Runtime Governance for AI Agents: Policies on Paths} \cite{kaptein2026paths}, which formalizes compliance policies as deterministic functions mapping agent identity, partial path, proposed next action, and organizational state to a violation probability. The two frameworks share the central insight that the runtime is the right object of governance and that prompts and static access control are insufficient. They differ in unit of analysis and emphasis.

Path-policy governance treats the execution path as the central object: a policy is a function over the partial path so far. The framework is most precise about \emph{when} a policy can fire (anywhere along a path) and about the relationship between path-dependent policies and design-time controls. SARC takes the constraint as the central object and asks where in the agent loop each constraint class is enforceable. The frameworks are largely complementary: a SARC enforcement point is the runtime location at which a path-policy function would be evaluated, and SARC's class-determined placement (hard at PAG, soft at PAA, escalation through ER) corresponds to a typing of the path-policy function space by the kind of decision the policy renders. SARC adds: a constraint-class taxonomy, the reversibility-window-aware escalation queue, the EU AI Act compilation path with Annex III mapping, the eight invariants for specification-trace correspondence, and the multi-agent extension with attribution-preserving trace trees. Path-policy governance adds: a precise probabilistic formulation of policy violation likelihood and a clean account of the path-dependence that prompts and access control cannot capture. A practical implementation of SARC would naturally use path-policy semantics for predicate evaluation at PAG, ATM, and PAA, and a path-policy implementation would naturally adopt SARC-style constraint-class typing for the policies it evaluates. We treat the two as members of the same emerging research program, distinguished primarily by what they make first-class rather than by what they exclude.

\subsection{Comparison with Adjacent Frameworks}\label{sec:comparison}

Table~\ref{tab:comparison} situates SARC against the closest adjacent frameworks. The point of the table is not to claim superiority but to make the division of labor explicit.

\begin{table}[h]
\centering\footnotesize
\begin{tabularx}{\textwidth}{@{}lXXX@{}}
\toprule
\textbf{Framework} & \textbf{Stage} & \textbf{Object} & \textbf{Constraint treatment} \\
\midrule
MDP \cite{puterman1994mdp} & Optimization & $(S,A,P,R,\gamma)$ & Implicit in $P$ or $R$ \\
CMDP \cite{altman1999cmdp} & Optimization & MDP + cost vector $\mathbf{d}$, bounds $\bar{\mathbf{d}}$ & Explicit cost vector with admissibility bounds \\
Safe RL \cite{garcia2015safe,achiam2017cpo} & Algorithm & Constrained policy optimization & Algorithmic enforcement during training/deployment \\
Policy-as-code \cite{jin2017opa,baier2024policymachine} & Implementation & Generic policy decision engine & Substrate for evaluating arbitrary policy predicates \\
Runtime enforcement \cite{schneider2003enforcement,fournet2003stack} & Runtime & Reference monitor / security automaton & Generic mechanism for intercepting policy-relevant events \\
\textbf{SARC} (this paper) & \textbf{Specification \& Architecture} & $\langle S,A,R,C\rangle$ + four enforcement points & First-class constraint primitive, agent-loop-specific placement, regulatory compilation \\
\bottomrule
\end{tabularx}
\caption{SARC and adjacent frameworks. SARC sits upstream of optimization frameworks (MDP, CMDP, safe RL) and is implementable on top of generic substrates (policy-as-code, runtime enforcement). What SARC contributes that none of the others does is an agent-specific specification of \emph{which} constraint goes \emph{where} in the agent loop, with regulatory compilation as a first-class concern.}
\label{tab:comparison}
\end{table}

\subsection{What SARC Is Not}\label{sec:not}

To prevent the framework from accumulating unwarranted scope through reader inference, we are explicit about what SARC does not provide.

\begin{itemize}[leftmargin=2em,topsep=0.3em,itemsep=0.3em]
\item SARC is \emph{not} a legal compliance certification. A SARC-conformant deployment is a system whose runtime is auditable against its declared specification. Whether the specification correctly encodes a regulator's interpretation of an obligation is an institutional question SARC does not resolve.
\item SARC is \emph{not} a safety guarantee. The framework prevents inadmissible actions from executing only to the extent that the predicates correctly capture inadmissibility. A predicate that misses a class of harmful actions produces a system that is auditable against the wrong specification.
\item SARC is \emph{not} a learning algorithm. It does not specify how a policy is trained, how rewards are shaped, or how exploration is bounded. Safe-RL methods compose with SARC; they do not substitute for it.
\item SARC is \emph{not} a prompt-engineering discipline. Constraint-aware planning is an efficiency mechanism (\S\ref{sec:loop}) that operates upstream of the enforcement surface. A system that relies on prompt-level constraint awareness without the runtime enforcement architecture is not SARC-conformant.
\item SARC is \emph{not} an evaluation benchmark. The empirical evaluation in \S\ref{sec:sim} is a controlled comparison against practitioner baselines on a synthetic procurement task; it characterizes the structural and measurable consequences of placement-and-typing discipline. It is not a benchmark suite for the broader question of agent governance, and the follow-up work scoped in \S\ref{sec:future-impl} (LangGraph integration, AgentDojo adversarial evaluation, additional task domains, real-deployment shadow rollout) is the natural empirical extension.
\end{itemize}

The framework is what is left when these are separated from it: a specification format, a runtime architecture, an audit-checking property, and a multi-agent extension. Its value is in what it makes possible (the institutional, learning, prompting, and evaluation work above), not in substituting for any of them.

\section{The SARC Construct}\label{sec:def}

\subsection{Definition}

\begin{definition}[Agentic System Specification]
An agentic system specification is a tuple
\[
    \Sigma = \langle S, A, R, C \rangle
\]
where the first three components follow the standard RL semantics, with elaborations for the agentic-LLM setting, and the fourth is defined below.
\end{definition}

\paragraph{State $S$.} $S$ denotes the set of representations the agent can observe and use to condition action selection. We adopt the partially observable view as default \cite{kaelbling1998pomdp}. Specification of $S$ requires declaring (i) the input modalities the agent ingests, (ii) the retrieval surfaces that hydrate context, (iii) the memory mechanism (stateless, episodic, persistent), and (iv) the freshness bounds beyond which observations are treated as stale.

\paragraph{Action Space $A$.} $A$ denotes the set of operations the agent may invoke. Specification of $A$ requires declaring (i) the registered tools and their typed signatures \cite{patil2024gorilla}, (ii) the maximal cardinality of action sequences before forced termination, and (iii) the cost function over individual actions, including invocation cost (compute, latency) and external cost (API spend, side effects on the world). We follow the agent-loop literature in distinguishing action selection from action plan, and require both to be bounded.

\paragraph{Reward $R$.} $R$ denotes the signal that drives action selection. We use the term in a deliberately broad sense: in many deployed LLM-agent systems, $R$ is not implemented as a learned reinforcement-learning reward signal but as a declared business objective, scoring rule, preference ordering, evaluation function, or heuristic ranking criterion. The framework is agnostic about the implementation; what it requires is that the architect declare what is being maximized and how. In production agentic systems $R$ is rarely a single scalar. We require $R$ to be specified as either (i) a scalarization of declared sub-rewards with explicit weights, or (ii) a lexicographic ordering with declared tie-breaking. Specification of $R$ requires declaring the time horizon, the asymmetry coefficient between false-positive and false-negative outcomes, and the credit assignment rule. Goodhart's principle \cite{manheim2019goodhart} is treated as a binding constraint on $R$ specification: any proxy reward must be paired with an integrity check.

\paragraph{Constraints $C$: the contribution.} $C$ denotes the set of bounds on agent behavior that must hold independently of reward maximization. Formally,
\[
    C = C_h \,\cup\, C_s \,\cup\, C_e
\]
partitioned into three classes:
\begin{itemize}[leftmargin=2em,topsep=0.3em,itemsep=0.2em]
\item $C_h$ \textbf{(hard constraints)}: violations are not permitted under any policy.
\item $C_s$ \textbf{(soft constraints)}: violations are admissible at a declared marginal cost.
\item $C_e$ \textbf{(escalation constraints)}: conditions under which control must be transferred to a human operator.
\end{itemize}
Each constraint $c \in C$ is a quintuple
\[
    c = \langle \texttt{src}, \texttt{class}, \texttt{pred}, \texttt{verif}, \texttt{resp} \rangle
\]
where \texttt{src} is the source (regulatory, contractual, ethical, operational); \texttt{class} $\in \{h, s, e\}$; \texttt{pred} is a predicate over $(s, a, t)$; \texttt{verif} is the verification mechanism (precondition check, runtime monitor, post-hoc audit); and \texttt{resp} is the response protocol (block, log, escalate, abort). $C$ subsumes the kill switch, the human-in-the-loop trigger, and the conformity surface, and renders them machine-readable.

\subsection{Two Scoping Propositions}

\begin{proposition}[Constraint Primacy under Finite Reward Shaping]\label{prop:primacy}
There exist deployment settings, common in regulated environments, in which no finite additive penalty term added to a base reward function can guarantee that every reward-optimal policy satisfies a hard constraint $c \in C_h$, while preserving the base optimization objective. Equivalently, constraint specification cannot, in general, be replaced by finite reward shaping when hard constraints correspond to low-probability, high-payoff violations.
\end{proposition}

A formal counterexample is constructed in \S\ref{sec:counter}. The proposition is intentionally narrow: it claims existence of failure cases, not that reward shaping is uniformly inadequate.

\begin{proposition}[Conditional Sufficiency]\label{prop:suff}
If $S$, $A$, $R$, and $C$ are each specified to the standards declared above, and if (i) the predicates in $C$ are decidable in the time available at their assigned enforcement point, and (ii) the escalation router has bounded queueing latency relative to the action's reversibility window, then $\Sigma$ admits a well-posed constrained decision problem reducible to a CMDP under standard measurability and feasibility assumptions \cite{altman1999cmdp}, and produces structured runtime records that can support technical-documentation and post-market-monitoring obligations under EU AI Act Articles 11 and 72, subject to the broader organizational and conformity-assessment requirements of the Act.
\end{proposition}

We deliberately weaken the earlier ``local sufficiency'' claim. Sufficiency is conditional on decidability and queue-capacity assumptions which are themselves engineering properties; without them, a SARC specification is well-formed but not necessarily executable. \S\ref{sec:ops} characterizes the conditions under which they hold.

\subsection{A Formal Counterexample to Reward-Only Encoding}\label{sec:counter}

We construct a minimal CMDP-like environment in which any finite reward function fails to encode a hard constraint that a CMDP cost-vector encodes trivially. The example is small but structurally generic.

\begin{example}[Reward shaping fails for hard constraints under finite penalty]\label{ex:counter}
Consider a two-state environment $S = \{s_0, s_1\}$ with a single decision at $s_0$ and actions $A = \{a_{\text{safe}}, a_{\text{risky}}\}$. Transitions: $a_{\text{safe}}$ deterministically remains in $s_0$ and yields reward $0$; $a_{\text{risky}}$ yields stochastic outcome, with probability $1{-}\epsilon$ a high reward $G > 0$ and remains in $s_0$; with probability $\epsilon$ a constraint violation that transitions to terminal state $s_1$. The hard constraint $c \in C_h$ states: \emph{violation must not occur}. Suppose we attempt to encode $c$ as a reward penalty $-M$ on the violation event, with $M < \infty$.

The expected return of $a_{\text{risky}}$ over a single step is
\[
    \mathbb{E}[R \mid a_{\text{risky}}] = (1-\epsilon) G - \epsilon M.
\]
The risky action is preferred to the safe action when $\mathbb{E}[R \mid a_{\text{risky}}] > \mathbb{E}[R \mid a_{\text{safe}}] = 0$, that is:
\begin{align*}
 (1-\epsilon) G - \epsilon M &> 0 \\
    G - \epsilon G - \epsilon M &> 0 \\
    G &> \epsilon (G + M) \\
    \epsilon &< \frac{G}{G + M}.
\end{align*}
For any choice of finite $M$, choose $\epsilon < G / (G + M)$. Then the reward-optimal policy selects $a_{\text{risky}}$, violating $c$ with probability $\epsilon > 0$ per visit. As $M \to \infty$, numerical conditioning of the reward signal degrades unboundedly, which is the operational reason the safe-RL literature \cite{altman1999cmdp,achiam2017cpo} addresses hard constraints through the CMDP cost vector rather than through reward shaping.

By contrast, the CMDP encoding (a separate cost function $d(s,a) = \mathbb{1}[\text{violation}]$ with admissibility bound $\bar d = 0$) directly forbids any policy with non-zero violation probability, independent of $G$. The CMDP solution is to set $\Pr(a_{\text{risky}}) = 0$; the reward-shaping ``solution'' yields the wrong policy for any finite $M$ and any sufficiently small $\epsilon$.
\end{example}

\paragraph{Scope of the counterexample.} The counterexample does not claim that no reward function can ever represent a constraint. Rather, it shows that finite penalty-based reward shaping cannot generally preserve both the original optimization objective and hard-constraint satisfaction. The failure arises because the regulatorily forbidden action may remain reward-optimal whenever the violation probability is sufficiently small relative to the payoff. This is precisely why SARC treats hard constraints as architectural predicates rather than as terms in the reward.

\begin{remark}
The example captures a recurring structural pattern rather than a general theorem. Whenever a hard constraint corresponds to a low-probability, high-reward event, finite reward shaping is insufficient in the sense demonstrated above: the policy gradient prefers the expected upside, and pushing $M$ toward the regime where it dominates degrades the numerical conditioning of the reward signal. This is precisely the structure of regulatory hard constraints in agentic deployment: violations are rare, the immediate optimization payoff for taking the constrained action is real, and the regulatory cost of violation is not numerically commensurate with the operational reward.
\end{remark}

\subsection{Boundary Condition}\label{sec:boundary}

SARC reduces to the classical $(S, A, R)$ tuple under three jointly necessary conditions: all actions are reversible at zero cost, no obligation external to the reward function applies, and the loss function is symmetric across error directions. We refer to this as Regime A. Outside Regime A (Regime B), the $C$ component is load-bearing in the sense of Proposition~\ref{prop:primacy}, and any specification omitting it is incomplete in operationally consequential ways. The framework is designed for Regime B.

\paragraph{Regime B is not binary.} Real deployments rarely sit purely in either regime. A single agent may operate over a heterogeneous action space in which some action classes are reversible and others are not, or in which regulatory obligations apply only to subsets of users, jurisdictions, or use cases. A customer-service copilot, for example, may handle low-stakes informational queries (Regime A) and benefit credit decisions (Regime B) within the same conversation. SARC applies selectively to the action classes that exhibit Regime B exposure: the constraint set $C$ governs only the affected action classes, and $C$ may be empty for the remainder. The framework's discipline is to make the per-action-class regime determination explicit at specification time rather than to force a single regime classification on the deployment as a whole. In practice this means $C$ is partitioned by action class, with each class declaring whether $C_h$, $C_s$, or $C_e$ apply.

\subsection{What a Minimal SARC-Conformant System Must Satisfy}\label{sec:invariants}

The framework as developed in the preceding subsections is descriptive: it defines what a SARC specification \emph{is}. We now make it normative: we define what a SARC-conformant system must satisfy. A system is SARC-conformant if and only if it honors the following invariants. To avoid confusion with legal compliance, we use ``SARC-conformant'' throughout the rest of the paper to mean conformity with the runtime invariants defined in this subsection, not conformity with any external regulation; a SARC-conformant system is one whose runtime can be mechanically audited against its declared specification, which is necessary but not sufficient for any particular legal compliance claim. The invariants below are not best practices; they are minimum admissible conditions. A system that violates any single invariant is not under-implementing SARC; it is implementing something else.

\begin{enumerate}[label={\textbf{I\arabic*}.},leftmargin=2.6em,topsep=0.3em,itemsep=0.4em]
\item \textbf{Pre-action admissibility.} Every action dispatched by the agent passes a pre-action gate that enforces all hard constraints and predictive escalations decidable from $(s, a)$ at PAG. Enforcement is binding at execution: no action reaches the tool layer without admissibility having been determined and the corresponding tool-layer and policy-layer constraints being active in the dispatch context. This invariant operationalizes the pre-action and enforced-execution contracts of \S\ref{sec:loop}.

\item \textbf{Constraint completeness.} Every constraint $c \in C$ declares all five fields of the quintuple $\langle \texttt{src}, \texttt{class}, \texttt{pred}, \texttt{verif}, \texttt{resp} \rangle$ (\S\ref{sec:def}) plus a declared operating point $\theta$ (\S\ref{sec:calib}). A constraint missing any field is not a constraint; it is a comment.

\item \textbf{Trace derivation.} Every execution emits a structured trace whose schema is derived from $\Sigma$. The trace contains, at minimum, the pre-state, the action, the post-state, the constraints evaluated and their outcomes, and the attribution tuple of \S\ref{sec:attribution}. The trace is generated; it is not reconstructed.

\item \textbf{Bounded escalation.} Every escalation constraint $c \in C_e$ declares a reversibility window $\tau_{\text{rev}}$, and the escalation router has bounded queueing latency relative to $\tau_{\text{rev}}$. When the bound is exceeded, the system fails safe (default-deny; \S\ref{sec:queue}). Escalation without a bound is not human oversight; it is deferred autonomy.

\item \textbf{Authority intersection.} For multi-agent workflows, every action executes under the intersection of authorities along its call chain (\S\ref{sec:multi-auth}). Authority is monotonically non-increasing under traversal. Empty intersections fail safe.

\item \textbf{Layer-class compatibility.} Every constraint is hosted at a layer compatible with its class: hard constraints at the orchestration, tool, or policy layer, never solely at the prompt layer (\S\ref{sec:arch}, \S4.3). A regulatory hard constraint hosted only at the prompt layer fails this invariant.

\item \textbf{Enforcement completeness.} All executable actions in the system traverse at least one enforcement point compatible with their constraint class. No execution path bypasses the enforcement surface. This invariant is structural rather than per-action: it is a property of the runtime topology, verified by inspection of the dispatch graph against $\Sigma$. Hidden code paths, out-of-band tool invocations, and direct calls that bypass the agent loop are violations regardless of whether any individual action they enable would have passed PAG.

\item \textbf{Specification-trace correspondence.} The audit trail produced by a SARC-conformant system can be machine-checked against its specification. Given $\Sigma$ and a trace $T$, the auditor can verify that every action in $T$ passed the constraints declared in $\Sigma$ at the points declared in $\Sigma$.
\end{enumerate}

The invariants are stated in service of a single property: \emph{specification-trace correspondence by construction}. A system that satisfies I1--I8 is one for which an auditor can answer the central question of regulated agentic deployment (``did the agent comply?'') by mechanical inspection of the trace against the specification, without recourse to the system's developers, its prompts, or its post-hoc reasoning. This is what we mean by ``auditable by construction.''

\paragraph{Scope of the auditability claim.} We use ``auditable by construction'' in a narrow technical sense: for every action recorded in the runtime trace, an auditor can determine which constraints were applicable, which were evaluated, what outcome they produced, and what response was taken. The phrase does not mean that the system is \emph{compliant} by construction, \emph{safe} by construction, or free from specification error. Compliance depends on whether the predicates correctly encode the underlying obligations; safety depends on whether the obligations are themselves adequate; correctness depends on whether the implementation matches the specification. SARC structures the runtime so that the audit question is mechanically answerable; it does not relieve the deploying organization of the upstream interpretive and engineering work.

\paragraph{Relationship to the reference monitor.} The classical reference monitor of Anderson \cite{anderson1972security} and Lampson \cite{lampson1971protection} is defined by three properties: \emph{completeness} (the monitor is invoked on every protected access), \emph{isolation} (the monitor cannot be tampered with or bypassed), and \emph{verifiability} (the monitor is small enough to be analyzed for correctness). SARC instantiates these classical reference monitor properties within the agent execution loop, with the eight invariants of \S\ref{sec:invariants} as the concrete instantiation. Table~\ref{tab:refmon} maps the correspondence.

\begin{table}[!htbp]
\centering\small
\begin{tabular}{@{}p{3.5cm}p{4cm}p{6.5cm}@{}}
\toprule
\textbf{Reference monitor property} & \textbf{SARC invariant} & \textbf{Operational content} \\
\midrule
Completeness & I7 (Enforcement completeness) & Every executable action traverses at least one enforcement point compatible with its constraint class; no execution path bypasses the surface. \\
Isolation & I6 (Layer-class compatibility) & Hard constraints are hosted at the orchestration, tool, or policy layer, never solely at the prompt layer; constraints are not subject to model reinterpretation. \\
Verifiability & I8 (Specification-trace correspondence) & The trace is machine-checkable against the specification; every action's constraint evaluation is reconstructible from the SARC tuple. \\
\bottomrule
\end{tabular}
\caption{Mapping between classical reference monitor properties and SARC invariants. The framework instantiates the three reference monitor requirements within the agent execution loop, with PAG, ATM, PAA, and ER serving as the concrete enforcement points.}
\label{tab:refmon}
\end{table}

\subsection{Specification-Trace Correspondence as a Checkable Property}\label{sec:audit-checker}

The auditability invariant I8 can be stated as a formal property of SARC-governed executions. We give the definition here so that ``auditable by construction'' has a precise computational meaning rather than a rhetorical one.

\begin{definition}[Specification-trace correspondence]\label{def:stc}
Let $\Sigma = \langle S, A, R, C \rangle$ be a SARC specification and let $T = \langle t_1, t_2, \ldots, t_n \rangle$ be a runtime trace, where each $t_i$ is a record of the form
\[
    t_i = \langle s_i, a_i, s'_i, r_i, \text{obs}_i, E_i, \alpha_i \rangle,
\]
with $E_i \subseteq C$ the set of constraints whose evaluation outcomes are recorded for action $a_i$, and $\alpha_i$ the attribution tuple. We say that $T$ is in correspondence with $\Sigma$, written $T \models \Sigma$, iff for every $t_i \in T$:
\begin{enumerate}[leftmargin=2em,topsep=0.2em,itemsep=0.2em]
\item \emph{(Coverage)} For every $c \in C$ such that $c$ applies to $a_i$ in state $s_i$, $c \in E_i$ or $c$ has been evaluated at the decidability-rescue layer $j'$ of \S\ref{sec:multi-prop} and the rescue is recorded.
\item \emph{(Class-placement compatibility)} For every $c \in E_i$, the verification point at which $c$ was evaluated is compatible with $\texttt{class}(c)$ as declared in $\Sigma$ (hard constraints not solely at the prompt layer; escalation constraints with bounded reversibility window; etc.).
\item \emph{(Outcome consistency)} For every $c \in E_i$ that fired, the recorded response in $t_i$ is the response declared by $c.\texttt{resp}$ in $\Sigma$.
\item \emph{(Attribution completeness)} The attribution tuple $\alpha_i$ resolves to a non-empty principal-and-agent chain whose intersected authority is non-empty.
\end{enumerate}
\end{definition}

\begin{property}[Decidable audit]\label{prop:decidable-audit}
For any finite $\Sigma$ and finite trace $T$, the question ``does $T \models \Sigma$ hold?'' is decidable in time $O(|T| \cdot |C|)$ given that each constraint's predicate is itself decidable in bounded time at its declared verification point. The check requires no access to the model that produced the trace or to its prompts.
\end{property}

The decidable-audit property is the formal content of ``auditable by construction.'' It states that an external auditor, given $\Sigma$ and $T$, can mechanically verify whether the system honored its specification, without requiring access to the model, the developers, or the deployment's institutional context. The check operates as a function over two artifacts that the framework requires the system to produce: the specification (declared at design time, versioned per \S\ref{sec:translation}) and the trace (emitted at runtime per Algorithm~\ref{alg:loop}).

\paragraph{What an audit checker does.} A checker implementing Definition~\ref{def:stc} performs four passes over $T$: (i) for each $t_i$, enumerate constraints in $\Sigma$ applicable to $a_i$ and check coverage against $E_i$; (ii) for each evaluated constraint, check that the verification point recorded in $t_i$ is compatible with the constraint's declared class; (iii) for each fired constraint, check that the response recorded in $t_i$ matches $c.\texttt{resp}$; (iv) for each $t_i$, check that $\alpha_i$ is well-formed and that the implied authority chain is non-empty. Failures at any pass produce a structured discrepancy report identifying the action, the constraint, and the violated condition. We have implemented this checker as a Python function, included as supplementary material at \url{https://github.com/besanson/sarc-governance}, operating over JSON specifications and runtime traces. The implementation is deterministic, predicate-language-agnostic, and bounded by tool latency rather than by model latency, in keeping with the framework's design rule that audit decisions should not depend on model inference. We exercise the checker on every SARC trace produced in the evaluation of \S\ref{sec:sim}.

\section{Reference Runtime Architecture}\label{sec:arch}

The contribution of SARC is not the four-tuple as a paper artifact; it is the four-tuple as an architectural commitment. This section makes the commitment concrete by specifying a reference runtime in which $C$ is enforced at four named points.

\subsection{The Four Enforcement Points}

\begin{enumerate}[leftmargin=2em,topsep=0.3em,itemsep=0.4em]
\item \textbf{Pre-Action Gate (PAG).} Evaluated before an action is dispatched. Predicate of the form $\phi(s, a)$. Used for hard constraints and predictive escalations decidable from state and contemplated action alone.

\item \textbf{Action-Time Monitor (ATM).} Evaluated during action execution, typically through wrapping of the tool call. Used for constraints that depend on partial outputs (cost accumulation, latency budget, content classification on streaming returns). Can interrupt mid-flight.

\item \textbf{Post-Action Auditor (PAA).} Evaluated after an action completes, before the next agent step. Used for constraints that compare post-action state to pre-action state (drift, side-effect verification). Cannot prevent the just-completed action; can prevent the next.

\item \textbf{Escalation Router (ER).} Stateless component invoked by the above. Routes a flagged action to the appropriate human operator, persists the decision, and returns control to the agent loop. \S\ref{sec:queue} characterizes its queueing behavior.
\end{enumerate}

\begin{figure}[h]
\centering
\begin{tikzpicture}[
    node distance=0.55cm and 0.4cm,
    layer/.style={draw=accent, fill=soft, rounded corners=2pt, minimum width=11.5cm, minimum height=0.85cm, font=\small\bfseries, align=center},
    enf/.style={draw=stop, fill=stop!10, rounded corners=2pt, font=\scriptsize\bfseries, align=center, minimum height=0.55cm, text=stop},
    flow/.style={->, thick, accent},
    tag/.style={font=\scriptsize\itshape, text=accent}
]

\node[layer] (orch) {Orchestration Layer \quad{\normalfont (planner $\cdot$ memory $\cdot$ tool dispatcher)}};
\node[layer, below=of orch] (tool) {Tool Layer \quad{\normalfont (typed tool registry $\cdot$ external systems)}};
\node[layer, below=of tool] (env)  {Environment / World \quad{\normalfont (databases $\cdot$ APIs $\cdot$ humans $\cdot$ physical assets)}};
\node[layer, above=of orch] (model) {Model Layer \quad{\normalfont (LLM $\cdot$ reasoning $\cdot$ planning prompts)}};
\node[layer, above=of model, fill=accent!15] (spec) {Specification $\Sigma = \langle S,A,R,C\rangle$};

\node[enf, right=0.4cm of orch.east, anchor=west, yshift=0.3cm] (pag) {PAG: Pre-Action Gate};
\node[enf, right=0.4cm of orch.east, anchor=west, yshift=-0.3cm] (atm) {ATM: Action-Time Monitor};
\node[enf, right=0.4cm of tool.east, anchor=west, yshift=0.0cm] (paa) {PAA: Post-Action Auditor};
\node[enf, right=0.4cm of model.east, anchor=west, yshift=0.0cm] (er)  {ER: Escalation Router};

\draw[flow, dashed] (spec) -- node[tag,right] {compiles to checks} (model);
\draw[flow] (model) -- (orch);
\draw[flow] (orch) -- (tool);
\draw[flow] (tool) -- (env);

\draw[->, thick, stop, dashed] (pag.west) -- ([yshift=0.3cm]orch.east);
\draw[->, thick, stop, dashed] (atm.west) -- ([yshift=-0.3cm]orch.east);
\draw[->, thick, stop, dashed] (paa.west) -- (tool.east);
\draw[->, thick, stop, dashed] (er.west)  -- (model.east);

\end{tikzpicture}
\caption{Reference runtime architecture. The specification $\Sigma$ compiles into runtime checks attached at four enforcement points. PAG sits between the planner's intended action and the tool dispatcher; ATM wraps tool execution; PAA inspects the post-action state before the next loop iteration; ER routes any flagged event to a human operator and returns the ruling.}
\label{fig:arch}
\end{figure}
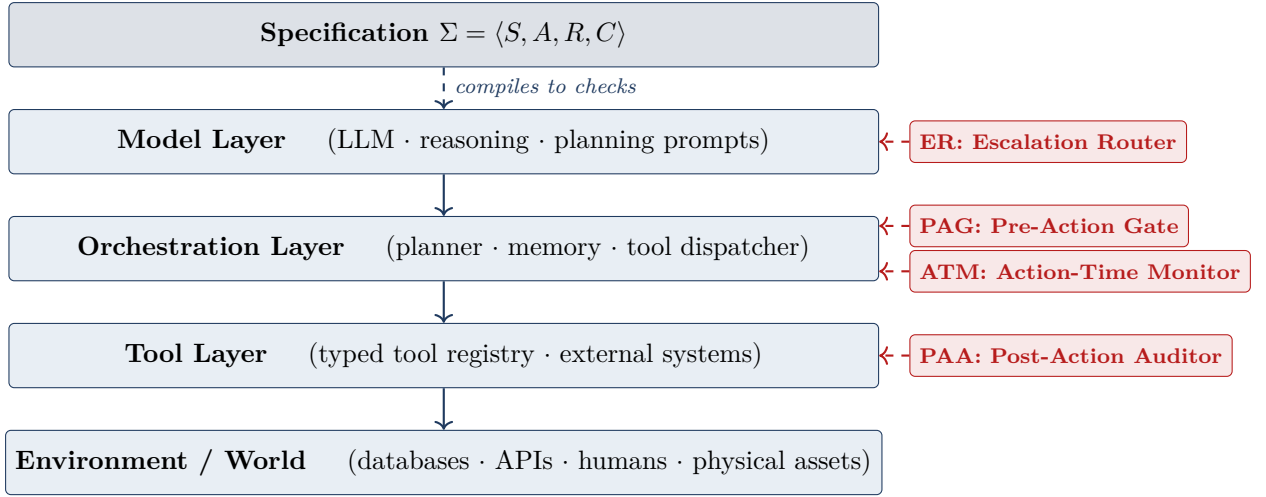

\subsection{Mapping Constraint Classes to Enforcement Points}

Each constraint class has a canonical enforcement point determined by its semantics, summarized in Table~\ref{tab:mapping}. Hard constraints belong at PAG (or ATM if the action is streaming) because admissibility must be determined before or during dispatch. Soft constraints belong at ATM or PAA where partial or completed action data is available for cost-weighted evaluation. Escalation constraints belong at PAG when predictive (firing on action signature alone) or PAA when reactive (firing on observed post-state).

\begin{table}[h]
\centering\small
\begin{tabularx}{\textwidth}{@{}lXXX@{}}
\toprule
\textbf{Constraint class} & \textbf{Enforcement point} & \textbf{Verification semantics} & \textbf{Failure response} \\
\midrule
Hard ($C_h$) & PAG; ATM if action is streaming & Decidable from $(s,a)$ before or during dispatch & Block; abort \\
Soft ($C_s$) & ATM; PAA & Decidable on partial or completed action; cost-weighted & Log; throttle; emit signal to $R$ \\
Escalation ($C_e$) & PAG (predictive); PAA (reactive) & Threshold predicate over $(s, a)$ or post-state & Suspend; route to ER; await ruling \\
\bottomrule
\end{tabularx}
\caption{Canonical mapping from constraint class to enforcement point. The strongest applicable point is preferred: a hard constraint that can be evaluated at PAG should never be deferred to PAA, where the action has already executed.}
\label{tab:mapping}
\end{table}

\subsection{The Layer Question: Where Does $C$ Live?}

A practical question for any team building under SARC is which architectural layer hosts the constraint logic. There are four candidate layers, in ascending order of robustness:

\begin{itemize}[leftmargin=2em,topsep=0.3em,itemsep=0.3em]
\item \textbf{Prompt layer.} Constraints expressed as natural-language instructions. Robustness: weak. Acceptable for low-stakes preferences; unacceptable for $C_h$.
\item \textbf{Orchestration layer.} Constraints expressed as code in the planner or dispatcher. Robustness: medium. Bypassable if the agent finds an alternate path.
\item \textbf{Tool layer.} Constraints embedded in the tool implementations themselves. Robustness: strong. The agent cannot route around a tool that refuses.
\item \textbf{Policy layer.} Constraints enforced outside the agent process entirely (API gateways, network policies, IAM rules, regulated infrastructure) \cite{jin2017opa,baier2024policymachine}. Robustness: strongest.
\end{itemize}

The discipline SARC imposes is to host each constraint at the lowest layer compatible with its class. A regulatory hard constraint hosted only at the prompt layer is not a constraint; it is an aspiration.

\section{Operational Trade-offs: Latency, Calibration, Queueing}\label{sec:ops}

Constraint enforcement is not free. This section characterizes its operational cost, since the framework is only useful to the extent that those costs are explicit.

\subsection{Latency Budget}\label{sec:latency}

Every enforcement point introduces latency. Let $\ell_{\text{model}}$, $\ell_{\text{tool}}$, $\ell_{\text{PAG}}$, $\ell_{\text{ATM}}$, $\ell_{\text{PAA}}$ denote per-step latencies of the model call, the tool call, and the three local enforcement points respectively (the ER is queued and treated separately in \S\ref{sec:queue}). Total per-step latency is
\[
    \ell_{\text{step}} = \ell_{\text{model}} + \ell_{\text{PAG}} + \ell_{\text{tool}} + \ell_{\text{ATM}} + \ell_{\text{PAA}}.
\]
For typical agentic systems $\ell_{\text{model}}$ is tens of seconds for a strong LLM and $\ell_{\text{tool}}$ ranges from sub-second (in-process function) to seconds (external API). Local constraint checks $\ell_{\text{PAG}}, \ell_{\text{PAA}}$ are normally implementable as in-process predicate evaluations (microseconds to low milliseconds for tabular checks; tens of milliseconds for retrieval-augmented checks); $\ell_{\text{ATM}}$ adds proportional overhead to streaming tool calls.

The design rule that emerges is operationally important: \emph{constraint evaluation should be O($\ell_{\text{tool}}$) or smaller, never O($\ell_{\text{model}}$)}. A check whose evaluation requires another LLM call is structurally suspect, both because it doubles latency per step and because it inherits the same nondeterminism the check is supposed to bound. Predicates over typed signatures, value ranges, identity, retrieval-source provenance, and structured tool outputs are appropriate; predicates that ask a separate LLM ``is this action OK?'' are not. Where natural-language reasoning is unavoidable in a check, it should be hosted as a distillation into a deterministic classifier or rule, not as a runtime LLM call.

\subsection{False-Positive / False-Negative Calibration}\label{sec:calib}

Each predicate $\phi$ has both error types. False positives (FP) block valid actions; false negatives (FN) allow violations. Treating each predicate as a binary classifier with operating point $\theta$ on a score, FP and FN are governed by the score distribution conditional on the violation indicator. The right operating point is a function of constraint class:

\begin{itemize}[leftmargin=2em,topsep=0.3em,itemsep=0.3em]
\item For $C_h$ (hard), the asymmetry favors FP: a missed violation is by definition unacceptable, while a blocked valid action falls back to escalation. The operating point is conservative; FP rate is paid.
\item For $C_s$ (soft), the asymmetry is symmetric in expectation: the predicate's purpose is to attach cost, not to block. The operating point is calibrated to match the empirical cost of the underlying violation.
\item For $C_e$ (escalation), the operating point is governed by operator capacity (\S\ref{sec:queue}): an escalation is a load on the human, and over-escalation degrades operator reliability through alert fatigue \cite{anaby2024alertfatigue,parasuraman1997vigilance}.
\end{itemize}

The architectural commitment is that operating points are declared in the specification, not selected ad hoc by implementers. A predicate without a declared operating point is incomplete.

\subsection{Escalation Queueing}\label{sec:queue}

Article 14 of the EU AI Act requires effective human oversight, which the framework operationalizes through the ER. ``Effective'' implies that operators are available when escalations arrive. Treating escalation arrivals as a Poisson process with rate $\lambda_e$ (per agent, per unit time) and operator service times as exponential with rate $\mu$, an M/M/c queue with $c$ operators yields expected wait
\[
    W_q = \frac{P_{\text{wait}}(c, \rho)}{c\mu - \lambda_e}, \quad \rho = \lambda_e/(c\mu),
\]
where $P_{\text{wait}}(c, \rho)$ denotes the Erlang-C probability that an arriving escalation must wait before being served, and $\rho$ is system utilization. Two operationally important consequences follow.

First, $\lambda_e$ is determined by the architect's choice of escalation predicate. Tightening $\phi$ for $C_e$ raises $\lambda_e$ and degrades $W_q$ super-linearly as $\rho \to 1$. The framework therefore couples constraint specification and operator staffing: an escalation predicate cannot be set independently of the staffing model that supports it.

Second, when $W_q$ exceeds the action's reversibility window (the time within which the contemplated action can be undone without compounding cost) the escalation is structurally too slow. The architect must then either tighten the predicate (escalate earlier), strengthen the constraint to $C_h$ (block instead of escalate), or accept that the system cannot be operated within Article 14's substance. The ``human is in the loop'' is not a statement; it is a measurable property of the queueing system.

\paragraph{Queue stability and operational unacceptability.} When $\rho \geq 1$, the escalation queue is unstable in the M/M/c sense: the expected wait diverges. Even when $\rho < 1$, escalation is operationally ineffective if the expected wait $W_q$ exceeds the action's reversibility window $\tau_{\text{rev}}$. In that case the system must either increase operator capacity, tighten the escalation predicate, convert the action class to a hard block, or default-deny. SARC requires a declared default-deny when no operator can service an escalation within $\tau_{\text{rev}}$: the action is treated as denied. This is a fail-safe, not a fail-degraded, default. Default-allow under operator unavailability converts $C_e$ into a no-op under load, which is the opposite of what Art.~14 requires.

\section{Compiling Regulatory Obligations into Runtime Checks}\label{sec:compile}

This section traces the path from a clause in a regulatory instrument to an executable predicate. We work through the EU AI Act \cite{eu2024aiact}, with reference to the legal scholarship interpreting it \cite{veale2021demystifying,smuha2021euapproach,malgieri2024aiact,laurent2024aiact,novelli2024riskmgmt}. The framework is agnostic about the specific instrument; the same path applies to financial regulation, contractual obligation, or sectoral rules. We emphasize that the interpretive arrow (from statutory text to operational obligation) is contested in the legal literature, and that SARC presupposes rather than resolves that interpretation.

\subsection{Compilation Path}

We define the compilation as a four-step mapping:
\[
    \underbrace{\text{Article}}_{\text{statute}} \;\longrightarrow\; \underbrace{\text{Obligation}}_{\text{normative}} \;\longrightarrow\; \underbrace{\text{Constraint } c}_{\text{spec-level}} \;\longrightarrow\; \underbrace{\text{Runtime check}}_{\text{executable}}
\]
The first arrow is interpretive (and contested in the legal literature \cite{veale2021demystifying}); the second is architectural; the third is implementation. SARC's architectural step is the slot in the specification into which the obligation is placed.

\subsection{Worked Examples}

\paragraph{Article 14 (human oversight).} Encoded as $c \in C_e$ at PAG: $\phi_{\text{escalate}}(s,a)$ is the disjunction of declared escalation triggers. The PAG suspends, ER routes to operator, ruling persists. Audit trail emerges as side effect.

\paragraph{Article 15 (robustness, accuracy, cybersecurity).} Splits into a soft constraint at PAA (drift detection between pre/post state distributions) and a hard constraint at the tool layer (security boundary; agent cannot route around an authenticated tool that refuses).

\paragraph{Article 72 (post-market monitoring).} Encoded as an always-on logging constraint at PAA whose schema is the SARC specification: $S$ generates state-distribution telemetry, $A$ generates action-and-cost telemetry, $R$ generates reward-decomposition telemetry, $C$ generates constraint-event telemetry. The Art.~72 reporting surface is the specification, machine-rendered.

\subsection{End-to-End Compilation Example}\label{sec:e2e}

The preceding subsections sketch each compilation step in isolation. We now trace one obligation through the full arrow from statutory text to runtime constraint tuple, to make the path concrete.

\paragraph{Step 1: Statutory text.} EU AI Act Article 14(1) requires that high-risk AI systems be ``designed and developed in such a way \ldots that they can be effectively overseen by natural persons during the period in which the AI system is in use.'' Article 14(4) elaborates that human oversight measures must enable the natural person to ``decide, in any particular situation, not to use the high-risk AI system or to otherwise disregard, override or reverse the output of the system.''

\paragraph{Step 2: Operational interpretation.} The translation layer of \S\ref{sec:compile} produces an operational reading: for the procurement-approval deployment of \S\ref{sec:sim} and Appendix~A, ``effective oversight'' means that any purchase order above a declared materiality threshold is suspended in flight until a designated human operator either approves, denies, or modifies the action within a bounded time window, with the system defaulting to denial if no decision arrives in time. The materiality threshold (\texteuro{}50{,}000) and the time window (600 seconds) are deployment parameters declared by the deploying organization in consultation with legal counsel and the relevant business owner.

\paragraph{Step 3: Predicate over $(s, a, t)$.} The operational reading compiles into a decidable predicate evaluated at PAG before dispatch:
\[
    \phi_{14}(s, a, t) \;=\; \bigl[\,\texttt{tool}(a) = \texttt{erp.create\_po}\bigr] \,\wedge\, \bigl[\,\texttt{amount}(a) \geq 50000\,\bigr],
\]
where $\texttt{tool}(a)$ extracts the tool identifier of the contemplated action and $\texttt{amount}(a)$ extracts the monetary amount from the action's typed signature. The predicate is exact (no classifier error) and decidable in $O(1)$ over the action signature.

\paragraph{Step 4: Constraint tuple.} The complete SARC constraint object is
\[
\setlength{\arraycolsep}{0.4em}
\begin{array}{rl}
c_{14} \;=\; \bigl\langle & \texttt{src} = \text{Art.~14}, \\
  & \texttt{class} = C_e, \\
  & \texttt{pred} = \phi_{14}(s, a, t), \\
  & \texttt{verif} = \text{PAG}, \\
  & \texttt{resp} = \text{escalate-to-human}(\tau_{\text{rev}} = 600\text{s},\, \text{on-timeout} = \text{deny}) \,\bigr\rangle.
\end{array}
\]

\paragraph{Step 5: Runtime emission.} When $\phi_{14}$ fires at PAG, the agent loop (Algorithm~\ref{alg:loop}) routes the action to the Escalation Router, which dispatches it to the procurement-managers operator group declared in the YAML of Appendix~A. The router holds the action up to $\tau_{\text{rev}}$. If the operator approves, the action proceeds to dispatch under the enforced constraint context. If the operator denies, the agent re-plans. If the timeout elapses, the action is denied (default-deny). Every transition emits a structured trace record with attribution tuple $\alpha$ and constraint identifier $c_{14}$, generating exactly the audit surface Article~72 reporting requires.

\paragraph{What the example illustrates.} The compilation arrow is not metaphor. Each step is mechanical given the preceding step: the statutory text is fixed, the operational interpretation is a documented institutional decision, the predicate is a typed function over the action signature, the tuple is a five-field record, and the runtime emission follows from Algorithm~\ref{alg:loop}. A SARC-conformant deployment can produce, for any constraint in $C$, the same five-step trace from statute to predicate to runtime check, which is what makes the system auditable by construction in the sense of \S\ref{sec:invariants}.

\subsection{Annex III Mapping}\label{sec:annex3}

Annex III of the EU AI Act lists the high-risk system categories. Table~\ref{tab:annex3} maps each Annex III category to the constraint patterns SARC identifies as load-bearing for that domain. The mapping is not exhaustive but identifies the dominant pattern.

\begin{table}[h]
\centering\footnotesize
\begin{tabularx}{\textwidth}{@{}p{3.4cm}p{2.4cm}X@{}}
\toprule
\textbf{Annex III category} & \textbf{Dominant class} & \textbf{Representative constraint pattern} \\
\midrule
Biometric identification & $C_h$, $C_e$ & Hard: opt-out registry check; escalation: any first-time identification of a flagged subject. \\
Critical infrastructure & $C_h$, $C_e$ & Hard: action-rate ceiling; escalation: any action affecting redundancy posture. \\
Education \& vocational training & $C_e$ & Escalation: assessment decisions affecting student progression. \\
Employment, worker management & $C_h$, $C_e$ & Hard: protected-class predicates blocked from inputs; escalation: termination, hire, promotion. \\
Access to essential services & $C_h$, $C_e$ & Hard: explainability gate (no action without contestable rationale); escalation: denials. \\
Law enforcement & $C_h$, $C_e$ & Hard: chain-of-custody integrity; escalation: any custodial decision. \\
Migration, asylum, border control & $C_h$, $C_e$ & Hard: prohibited-purpose registry; escalation: any individual-level decision. \\
Administration of justice & $C_h$, $C_e$ & Hard: no autonomous decision; system advises, judge decides. Escalation by definition. \\
Democratic processes & $C_h$ & Hard: prohibited-content registry, distribution boundary at tool layer. \\
\bottomrule
\end{tabularx}
\caption{Annex III categories mapped to dominant SARC constraint patterns. The table identifies typical patterns, not a completeness claim. Real deployments will combine multiple patterns and add domain-specific constraints.}
\label{tab:annex3}
\end{table}

\paragraph{What SARC covers and what it does not.} SARC covers the technical-runtime surface of compliance: the points in execution where obligations become checks and audit records become emissions. It does not cover the upstream surface (data governance, training-data lineage, model evaluation prior to deployment) or the institutional surface (governance boards, conformity assessment processes, market surveillance authority interactions). For high-risk systems under the AI Act, SARC is necessary but not sufficient for compliance; it is the runtime substrate on which compliance instruments rest.

\paragraph{Not legal compliance certification.} We state this boundary explicitly because the language of ``compilation'' invites a stronger reading than the framework supports. A SARC-conformant deployment is one whose runtime behavior can be mechanically audited against its declared specification. Whether the specification correctly encodes a given regulator's interpretation of an obligation is an institutional question that depends on legal review, conformity assessment, and ultimately judicial or regulatory determination, none of which SARC performs. A system that emits clean traces against a legally incorrect specification is auditable but not compliant. The framework's value is in narrowing the unaudited surface of a deployment, not in adjudicating the underlying legal interpretation. The translation layer of \S\ref{sec:translation}, governed as a versioned control surface, is the institutional mechanism that connects the runtime specification to evolving legal interpretation; the runtime is the substrate, the translation is the bridge, and certification is the regulator's determination, not ours.

\paragraph{The translation layer.}\label{sec:translation} The compilation path described above treats the move from obligation to predicate as a single arrow. In practice this arrow is an institutional process rather than a technical step. Translating a clause in a regulation, a paragraph in a contract, or a statement of policy into a machine-evaluable predicate requires cross-functional collaboration between legal counsel, compliance officers, domain experts, and engineering. The translation produces predicates that are auditable in two senses: the predicate text must be defensible against the source obligation, and the choice of operating point must be defensible against the cost coefficients of \S\ref{sec:economic}. SARC presupposes the existence of this translation layer and the institutional capacity to maintain it across regulatory change, but does not prescribe its implementation. A SARC-conformant deployment is therefore an artifact of two complementary investments: the runtime substrate, which this paper specifies, and the translation process, which the deploying organization owns.

\paragraph{The translation layer as a governed control surface.} In production deployments, this translation layer should be governed as a versioned control surface. Each predicate, operating point, threshold, and response protocol carries an owner, an approval record, an effective date, and a change history. Cross-functional sign-off by legal, compliance, domain, and engineering stakeholders is not part of the SARC runtime, but it is the institutional mechanism by which the runtime specification remains aligned with evolving legal interpretation. Without this discipline, predicates drift silently from their source obligations: a constraint written against a 2024 reading of Article 14 may no longer track the same article's interpretation under post-2026 jurisprudence, and a system that emits clean traces against a stale predicate is producing audit records that no longer reflect compliance. Treating $\Sigma$ as a versioned, owned, and reviewed artifact is the institutional analogue of treating it as an architecturally first-class object at runtime.

\section{The SARC-Governed Agent Loop}\label{sec:loop}

\begin{algorithm}[!htbp]
\caption{SARC-Governed Agent Loop}
\label{alg:loop}
\begin{algorithmic}[1]
\Require Specification $\Sigma = \langle S, A, R, C \rangle$ with $C = C_h \cup C_s \cup C_e$
\Require Initial state $s_0$, horizon $H$, model $\pi$, tool registry $\mathcal{T}$
\Require Attribution context $\alpha$ for the principal-and-agent chain
\State $t \gets 0$; $s \gets s_0$; $\text{trace} \gets [\,]$
\While{$t < H$ \textbf{and not} \textsc{Terminate}$(s)$}
    \State $a \gets \pi(s)$ \Comment{Plan: model proposes next action}
    \State \textbf{label} \textsc{PagCheck}: \Comment{Re-entry point after ER modifies $a$}
    \ForAll{$c \in C_h$ such that \texttt{verif}$(c) = \text{PAG}$}
        \If{$c.\texttt{pred}(s, a) = \text{violated}$}
            \State \Return \textsc{Abort}$(\text{trace}, c)$ \Comment{Hard block}
        \EndIf
    \EndFor
    \ForAll{$c \in C_e$ such that \texttt{verif}$(c) = \text{PAG}$}
        \If{$c.\texttt{pred}(s, a) = \text{trigger}$}
            \State $\text{ruling} \gets \textsc{ER}(s, a, c, \tau_{\text{rev}})$ \Comment{$\tau_{\text{rev}}$ = reversibility window}
            \If{ruling = \textsc{Timeout}} \State \Return \textsc{Abort}$(\text{trace}, c)$ \Comment{Default-deny}
            \ElsIf{ruling = deny} \State \textbf{continue while} \Comment{Re-plan}
            \ElsIf{ruling \textbf{modifies action}}
                \State $a \gets \text{ruling}.a$
                \State \textbf{goto} \textsc{PagCheck} \Comment{Re-validate modified action}
            \EndIf
        \EndIf
    \EndFor
    \ForAll{$c \in C_h$ such that \texttt{verif}$(c) \in \{\text{tool\_layer}, \text{policy\_layer}\}$ \textbf{and} \textsc{Applies}$(c, a)$}
        \State \textbf{assert} \textsc{EnforcementActive}$(c, a)$ \Comment{Tool/policy-layer check is wired up}
    \EndFor
    \State $(s', \text{obs}) \gets \textsc{Dispatch}\bigl(a, \mathcal{T} \mid C_{\text{tool}}, C_{\text{policy}}\bigr)$ \textbf{under enforced constraint context}, with ATM wrapping
    \ForAll{$c \in C_s \cup C_e$ such that \texttt{verif}$(c) = \text{PAA}$}
        \If{$c.\texttt{pred}(s, s', a, \text{obs}) = \text{fire}$}
            \State \textsc{Respond}$(c, s, s', a)$
        \EndIf
    \EndFor
    \State $r \gets R(s, a, s')$
    \State $\text{trace}.\textsc{Append}\bigl(\textsc{EmitTrace}(\Sigma, s, a, s', r, \text{obs}, c\text{-events}, \alpha)\bigr)$ \Comment{I3, I8; Art.~72}
    \State $s \gets s'$; $t \gets t + 1$
\EndWhile
\State \Return $\text{trace}$
\end{algorithmic}
\end{algorithm}

The loop is unsurprising in shape; what makes it SARC-governed is the explicit bracketing of the action by constraint evaluation, the timeout-default-deny on the ER, the re-validation of any operator-modified action against PAG before dispatch, and the structural binding of execution to the active tool-layer and policy-layer constraints. The notation $\textsc{Dispatch}(a, \mathcal{T} \mid C_{\text{tool}}, C_{\text{policy}})$ is not assertional but executional: dispatch occurs in a constraint context within which $C_{\text{tool}}$ and $C_{\text{policy}}$ are active enforcement scopes, not predicates that have been checked and then released. An action that satisfies the PAG checks but for which the corresponding tool-layer or policy-layer enforcement is not active in the dispatch context does not execute. Three contracts are visible in the algorithm. The \emph{pre-action contract}: no action is dispatched without all PAG-classed checks evaluating to admissible, including after any ER modification. The \emph{enforced-execution contract}: dispatch is structurally bound to the active tool-layer and policy-layer constraint context; constraints are enforced at execution, not merely checked beforehand. The \emph{post-action contract}: no next iteration begins without all PAA-classed checks having been evaluated. The \emph{telemetry contract}: every iteration produces a structured record whose schema is the specification, and which carries the attribution tuple $\alpha$, jointly satisfying invariants I3 and I8 of \S\ref{sec:invariants}.

\paragraph{Constraint-aware planning is upstream of enforcement.} A naive planner that ignores $C$ produces actions that are repeatedly blocked at PAG, wasting model calls and operator-attention budget on the escalation queue. In practice, deployments operate planners that are constraint-aware: the prompt or the planning policy is conditioned on a summary of $C$, so the planner anticipates likely violations and adjusts its trajectory before dispatch. This is a legitimate and often necessary engineering optimization. It is not a substitute for enforcement, and it does not weaken Proposition~\ref{prop:primacy}. Planning-time constraint awareness reduces the rate at which violations are \emph{attempted}; it cannot guarantee the rate at which violations are \emph{executed}, because planning operates on the planner's belief about the world and not on the world itself. The planner can be wrong about whether a constraint applies, the predicate library at planning time may differ from the runtime predicate library, and the constraint state may have changed between plan and dispatch. The enforcement surface remains the source of runtime admissibility control; the planner's awareness of $C$ is an efficiency property, not a correctness one. SARC accommodates constraint-aware planning as a deployment choice without changing what the runtime is required to do.

\paragraph{Planning-to-queueing: the operational consequence.} Constraint-aware planning is therefore an efficiency mechanism, not a safety mechanism, and its operational value is that it reduces the rate of attempted violations before they reach PAG or ER. This directly lowers the escalation arrival rate $\lambda_e$, queue utilization $\rho = \lambda_e/(c\mu)$, and expected wait $W_q$ in the escalation model of \S\ref{sec:queue}. The design trade-off is planning complexity against operational load: richer constraint awareness increases planning cost (model calls per step, prompt length, planner latency) but reduces failed dispatch attempts, operator burden, and wait time under load. Under high load, this trade-off can dominate: a deployment whose escalation queue is saturated benefits more from a constraint-aware planner that halves $\lambda_e$ than from staffing additional operators. The framework's discipline is to treat the planner's constraint awareness as a tunable input to the queueing model rather than as a substitute for the enforcement surface; runtime enforcement remains the correctness boundary, and constraint-aware planning is the operational lever that keeps the queueing system inside the admissible region.

\section{Comparison with Existing Agent Governance Approaches}\label{sec:practitioner-compare}

The comparison in \S\ref{sec:comparison} situates SARC against academic decision-theoretic formalisms (MDP, CMDP, safe RL) and infrastructure-level abstractions (policy-as-code, runtime enforcement). Practitioners deploying agentic systems today rely on a different set of mechanisms, drawn from the AI tooling stack and from operational tradition. We compare against those in Table~\ref{tab:practitioner-compare}.

\begin{table}[h]
\centering\footnotesize
\begin{tabularx}{\textwidth}{@{}p{2.8cm}p{2.5cm}p{2.2cm}p{1.6cm}X@{}}
\toprule
\textbf{Approach} & \textbf{Where constraints live} & \textbf{When enforced} & \textbf{Auditability} & \textbf{Dominant failure mode} \\
\midrule
Prompt guardrails & Prompt layer (system prompt, instructions) & Pre-generation (hopeful) & Weak; depends on model compliance & Bypass via tool calls whose effects are off-screen \\
Output filtering & Wrapper around LLM output & Post-generation, pre-action & Partial; only output-visible violations caught & Too late for any action whose effect is the side effect of a tool \\
Workflow rules & Orchestration code (if/else gates around tool dispatch) & Selectively, where coded & Medium; depends on coverage discipline & Path bypass when agent finds an unguarded route through the action space \\
Dashboard / human review & Operator UI, post-execution & After the fact & Documentary, not preventive & Alert fatigue; rubber-stamping under load \\
\textbf{SARC} & Runtime, multi-point (PAG, ATM, PAA, ER) & Pre, during, post; class-determined & Strong; specification-derived trace & Explicit trade-offs (latency, FP rate, queue load) characterized at design time \\
\bottomrule
\end{tabularx}
\caption{SARC compared to practitioner agent-governance approaches in current use. The comparison is not that SARC is universally superior; the leftmost three rows describe legitimate light-touch mechanisms suitable for low-stakes deployments. The point is that the four practitioner approaches each suffer a structural failure mode that SARC's class-determined enforcement is specifically designed to address.}
\label{tab:practitioner-compare}
\end{table}

\paragraph{The thesis line.} Existing approaches attach constraints to artifacts: to the prompt, to the output, to the orchestration code, to the operator's screen. SARC attaches constraints to execution. The shift from artifact-attachment to execution-attachment is the operational content of the framework's discipline. An artifact-attached constraint depends on whatever agent or operator interprets the artifact correctly; an execution-attached constraint runs whether or not the agent is well-behaved.

\paragraph{Dominance statements, by class.} The comparison admits four sharp claims, each tied to a specific failure mode the alternative cannot address.

\begin{itemize}[leftmargin=2em,topsep=0.3em,itemsep=0.3em]
\item Prompt guardrails cannot enforce hard constraints by construction because the model may ignore, reinterpret, or route around them through tools whose effects are off-screen.
\item Output filtering cannot enforce constraints whose violation is realized through side effects on the world rather than through visible text in the model output.
\item Post-hoc audit cannot prevent violations; it can only document them. The claim of \S\ref{sec:counter} that the gap between SARC and audit is qualitative, not quantitative, follows directly.
\item Workflow rules are insufficient in open or dynamically expanding action spaces unless they are tied to a complete action registry and enforced at dispatch; a workflow rule that does not see the action cannot block it.
\end{itemize}

The claim is not that SARC is universally superior because it is heavier. The claim is that where constraints are load-bearing (Regime B of \S\ref{sec:boundary}), SARC is structurally better suited because it binds constraint class to enforcement point. The lighter mechanisms are appropriate substitutes only outside Regime B.

\paragraph{When the practitioner approaches are appropriate.} Prompt guardrails are appropriate for stylistic and tonal preferences ($C$ classes weaker than what SARC formalizes). Output filtering is appropriate when the action surface is genuinely confined to text generation. Workflow rules are appropriate when the action space is small and statically known. Dashboards are appropriate as supplementary visibility on top of programmatic enforcement, not as primary enforcement. The framework argues that these mechanisms are not load-bearing for $C_h$ in regulated deployment, and that practitioners frequently treat them as if they were.

\section{Multi-Agent Systems Under SARC}\label{sec:multi}

Single-agent SARC governs the loop of one planner-executor; multi-agent SARC governs the lattice of relationships among multiple planners, executors, tools that are themselves agents, and principals on whose behalf the lattice acts. Most enterprise agentic deployments are multi-agent, often by accident rather than by design \cite{wu2023autogen,park2023generative,hong2024metagpt,liu2024agentbench}. The single-agent framework as developed in \S\ref{sec:def}--\S\ref{sec:loop} does not scale to this setting without explicit treatment of topology, propagation, authority composition, and cross-agent failure modes. This section provides that treatment.

The expanded treatment is motivated by an observation about practice: under-governed multi-agent systems exhibit failure modes that have no analogue in the single-agent case. A workflow that satisfies SARC at every individual agent can still violate the deploying organization's intent if constraints fail to propagate across an agent boundary, if authority is silently inherited from a more privileged principal, if attribution is summarized rather than preserved, or if a constraint laundering path exists between two agents with overlapping action spaces. We name these failure modes (\S\ref{sec:multi-failure}) and tie them back to architectural commitments in the runtime contract.

\subsection{Topologies}\label{sec:multi-topo}

Multi-agent systems differ structurally, and constraint propagation differs with them. We distinguish five canonical topologies; real deployments combine them.

\begin{enumerate}[leftmargin=2em,topsep=0.3em,itemsep=0.4em]
\item \textbf{Hierarchical (orchestrator-worker).} One agent decomposes a task and dispatches sub-tasks to specialized workers \cite{wu2023autogen}. Constraint propagation is downward: workers inherit from the orchestrator. Most studied; least general.
\item \textbf{Hierarchical, recursive (orchestrator-of-orchestrators).} A worker is itself an orchestrator dispatching further sub-agents \cite{hong2024metagpt}. Constraint propagation is recursive; depth is unbounded a priori.
\item \textbf{Pipeline.} Agents arranged in a directed acyclic chain, each consuming the previous agent's output. Constraints propagate forward; intermediate agents may not see the principal's full intent.
\item \textbf{Peer.} Agents communicate as equals; no single agent owns the workflow. Constraint propagation requires explicit handshake; there is no central locus to evaluate $C^*$.
\item \textbf{Market or auction.} Agents bid on tasks under a coordination protocol \cite{li2024marketauction}. Constraints govern the protocol itself rather than individual agents; bidders may be drawn from outside the deploying organization's trust boundary.
\end{enumerate}

The framework's coverage varies by topology. Hierarchical and pipeline topologies admit a clean orchestrator construct, formalized in \S\ref{sec:orch-loop}. Peer topologies require a virtual orchestrator: a stateless component that holds $C^*$ and is consulted before any cross-agent action. Market and auction topologies require constraint-on-protocol rather than constraint-on-agent and are partially treated in this paper; the harder question of how SARC governs a market in which counterparty agents are out of trust scope remains open.

\subsection{Constraint Propagation under Recursion}\label{sec:multi-prop}

The core technical question of multi-agent SARC is what happens to a constraint $c$ declared at the orchestrator when it is evaluated at depth $k$ in a recursive call chain. Three things can fail.

\paragraph{Decidability loss.} The predicate $c.\texttt{pred}(s, a)$ may be defined over orchestrator state $s$ that is not present in the depth-$k$ sub-agent's state space. A constraint requiring access to the principal's risk profile cannot be evaluated by a sub-agent that has been given only a tool signature. The framework's commitment is that an inherited constraint that cannot be evaluated at the layer where it must fire is not silently dropped; it is evaluated at the deepest layer (closest to execution) at which the constraint remains decidable, formalized below.

\paragraph{Decidability rescue, formally.} For a constraint $c$ declared at layer $i$ and reached at layer $j > i$ via a chain of agent invocations, define the decidability-rescue layer
\[
    j' \;\triangleq\; \max\bigl\{\, k \in [i, j] \,:\, c \text{ is decidable at layer } k \,\bigr\}.
\]
The set is non-empty because $c$ is decidable at its declaration layer $i$ by assumption; the maximum is therefore well-defined and lies in $[i, j]$. Constraint evaluation occurs at layer $j'$ rather than at layer $j$ when $j' < j$. The rationale for $\max$ rather than $\min$ is operational: evaluating at the deepest decidable layer keeps the evaluation as close to the action as possible, minimizing the staleness of the state the predicate observes and preserving the latency-budget bound of the next paragraph. The orchestrator-layer rescue described informally above is the special case $j' = i$, which arises when no intermediate layer has the state required to evaluate $c$.

\paragraph{Latency budget exhaustion.} Each enforcement point has a latency budget (\S\ref{sec:latency}). Inherited constraints aggregate: a depth-3 sub-agent inheriting twelve constraints from three layers above must evaluate them all within its own budget. The framework requires that latency budgets be declared per layer rather than per agent; an inherited constraint carries its declared latency budget down with it, and a sub-agent that cannot meet that budget refuses the call rather than degrades the check.

\paragraph{Authority degradation.} An action authorized at depth 0 may pass through agents whose individual authorities are narrower than the principal's. A naive inheritance would have the depth-$k$ executor act under the principal's full authority by transitivity, which collapses the authority surface. The framework requires that authority at each layer be the \emph{intersection} of the principal's authority and the layer's own authority, so that authority can be progressively narrowed but never broadened by traversal.

We summarize these as the propagation contract:

\begin{quote}\itshape
For every constraint $c$ declared at layer $i$ and reached at layer $j > i$ via a chain of agent invocations: (1) $c$ is evaluated at the decidability-rescue layer $j'$ defined above, or escalated when no decidable layer admits the budget; (2) $c$ carries its declared latency budget across layers, and a layer that cannot meet the budget refuses the call; (3) the authority under which $c$ is evaluated is the intersection of all per-layer authorities $i, i+1, \ldots, j'$ along the call chain.
\end{quote}

\subsection{Authority Composition Across Principals}\label{sec:multi-auth}

A multi-agent workflow can serve more than one principal. A procurement workflow involves a requester, a manager, finance, the supplier; a clinical decision-support workflow involves the patient, the clinician, the institution. The single-agent attribution tuple of \S\ref{sec:attribution} has one principal slot. The multi-agent extension requires that authority be composed across principals.

We distinguish two composition rules:

\begin{itemize}[leftmargin=2em,topsep=0.3em,itemsep=0.3em]
\item \textbf{All-of authority ($\bigcap$).} The action requires all relevant principals' authorities to be present and non-revoked. Used when each principal has a veto. A purchase requiring requester authorization \emph{and} manager approval \emph{and} finance budget release composes as $\bigcap$.
\item \textbf{Any-of authority ($\bigcup$).} The action requires at least one relevant principal's authority. Used when any qualified principal suffices. Emergency override patterns (any senior clinician can authorize an exception) compose as $\bigcup$.
\end{itemize}

The architect's task at specification time is to declare which composition applies to each action class. Silent default-to-any-of, in particular, is unsafe: an agentic system that proceeds when any of the relevant principals has authority will execute actions whose authority chain is incomplete. The framework's default is all-of; any-of must be declared explicitly with its qualifying-principal predicate.

\paragraph{Authority and constraint evaluation are coupled.} A constraint sourced from a contractual obligation between principal $A$ and principal $B$ cannot be evaluated against an action authorized only under $A$'s authority; the contract is bilateral. The framework requires that constraints carry an authority-binding field declaring which principals' authorities must be present for the constraint to be a check rather than a no-op. A regulatory constraint typically binds to all principals; an operational constraint typically binds to its source principal.

\subsection{The Orchestrator Loop}\label{sec:orch-loop}

The single-agent loop of Algorithm~\ref{alg:loop} governs one planner-executor. The orchestrator loop governs dispatch across agents. We give it as Algorithm~\ref{alg:orch}.

\begin{algorithm}[!htbp]
\caption{SARC-Governed Orchestrator Loop}
\label{alg:orch}
\begin{algorithmic}[1]
\Require Workflow specification $\Sigma^* = \langle S^*, A^*, R^*, C^* \rangle$ where $C^* = \bigcup_i C_i$
\Require Set of worker agents $\mathcal{W}$ with their specifications $\{\Sigma_w\}$
\Require Initial principal-and-agent chain $\mathbf{P} = [p_0]$, current dispatch node $\alpha_\tau$
\State Initialize trace tree; root attribution $\alpha_0 \gets \langle \mathbf{P}, \text{orch}, \text{orch}, \text{none}, \text{authority}(p_0), \emptyset \rangle$
\State $\alpha_\tau \gets \alpha_0$ \Comment{Current dispatch node starts at root}
\While{workflow incomplete}
    \State Plan: orchestrator selects sub-task $\tau$ and target worker $w \in \mathcal{W}$
    \State Construct dispatch context: state slice $s_w$, constraint subset $C^*_{|w}$ decidable in $s_w$
    \ForAll{$c \in C^*$ \textbf{not} decidable in $s_w$}
        \If{$c$ is hard or escalation}
            \State Evaluate $c$ at orchestrator layer before dispatch \Comment{Decidability rescue}
        \Else
            \State Defer to PAA at orchestrator after worker returns
        \EndIf
    \EndFor
    \State $\text{principal\_auth} \gets \textsc{ComposeAuthority}\bigl(\text{action\_class}(\tau), \textsc{Principals}(\tau), \text{rule} \in \{\text{all\_of}, \text{any\_of}\}\bigr)$
    \State $\text{chain\_auth} \gets \bigcap_{p \in \mathbf{P}} \text{authority}(p)$
    \State $\text{auth}_w \gets \text{principal\_auth} \,\cap\, \text{chain\_auth} \,\cap\, \text{authority}(w)$
    \If{$\text{auth}_w = \emptyset$}
        \State \textsc{Abort}(workflow, ``authority degradation to empty'') \Comment{See \S\ref{sec:multi-failure}}
    \EndIf
    \State Create dispatch node $\alpha_w$ as child of $\alpha_\tau$; record $\langle \mathbf{P} + [w], \tau, w, \text{auth}_w \rangle$
    \State Dispatch to worker with $\langle s_w, C^*_{|w}, \text{auth}_w, \mathbf{P} + [w], \alpha_w \rangle$
    \State Worker executes its own SARC-governed loop (Algorithm~\ref{alg:loop}); for nested orchestration the worker is itself an orchestrator and recursively invokes Algorithm~\ref{alg:orch} with $\alpha_\tau \gets \alpha_w$
    \State Receive worker trace $T_w$ with attribution tuples for every executed action
    \State \textbf{Fold:} append $T_w$ to the trace tree under the current dispatch node $\alpha_w$, preserving worker attribution; do not summarize
    \State Evaluate orchestrator-layer PAA constraints over the post-worker state
\EndWhile
\State \Return trace tree \Comment{Hierarchical, attribution-preserving, recursion-safe}
\end{algorithmic}
\end{algorithm}

Three properties of the orchestrator loop deserve emphasis. First, the trace is a tree, not a sequence. A multi-agent workflow's audit surface is necessarily hierarchical because the dispatch structure is hierarchical; flattening to a sequence loses the structural information a regulator needs. The fold attaches each worker's subtree to the current dispatch node $\alpha_w$, not to the root, so recursive workflows preserve their nesting structure rather than collapsing into the root. Second, the orchestrator is the locus of decidability rescue: constraints inherited from the workflow that cannot be evaluated at the worker's layer must be evaluated somewhere, and the orchestrator is that somewhere. Third, authority composition is two-stage: \textsc{ComposeAuthority} at the principal level (handling the all-of vs.\ any-of distinction of \S\ref{sec:multi-auth}), then intersection with the call-chain authority and the worker's own authority. The authority-empty check is a hard fail-safe: an empty intersection means the workflow as designed cannot legally proceed under the present principal chain, and the framework refuses to silently proceed.

\paragraph{Authority monotonicity, formally.} For any call chain $\mathbf{P} = [p_0, p_1, \ldots, p_k]$, define depth-$k$ authority as
\[
    \text{auth}_k \;\triangleq\; \bigcap_{i=0}^{k} \text{authority}(p_i).
\]
Adding any agent at depth $k+1$ yields $\text{auth}_{k+1} = \text{auth}_k \cap \text{authority}(p_{k+1}) \subseteq \text{auth}_k$. Authority is therefore monotonically non-increasing under traversal: it can narrow through delegation but cannot expand. Multi-principal composition (line 13) operates at a single depth and does not affect this property; it determines which principals' authorities feed into $\text{auth}_0$, after which monotonicity governs propagation downward.

\subsection{Cross-Agent Failure Modes}\label{sec:multi-failure}

The multi-agent setting introduces failure modes with no single-agent analogue. We name four.

\paragraph{Constraint laundering.} An agent passes work to a less-constrained agent in order to bypass a hard predicate. The classic instance: agent $\Sigma_1$ has a hard constraint forbidding action class $X$; agent $\Sigma_2$, invoked as a tool by $\Sigma_1$, has no such constraint and executes $X$ on $\Sigma_1$'s behalf. The framework's defense is propagation: the inheritance rule of \S\ref{sec:multi-prop} requires $\Sigma_2$ to inherit $\Sigma_1$'s hard constraint where decidable, with decidability rescue at $\Sigma_1$ where not. A worker that does not enforce the orchestrator's hard constraints is a laundering path by construction.

\paragraph{Authority escalation via tool capability.} A sub-agent's tool has broader scope than the calling agent's authority. The sub-agent invokes the tool successfully because the tool authenticates against the sub-agent's own credentials, not the calling chain's authorities. The framework's defense is the authority-intersection rule of \S\ref{sec:multi-auth}: tool calls execute under the intersected authority of the call chain, not the sub-agent's local authority. Tool implementations must respect the supplied authority context, not their own.

\paragraph{Trust-boundary violation.} A sub-agent imports state (retrieved documents, tool outputs, agent responses) from outside the principal's trust scope, contaminating the orchestrator's state. The classic instance is prompt injection across agent boundaries \cite{greshake2023injection,debenedetti2024agentdojo}: a worker retrieves a document containing adversarial instructions, the worker's response includes the injected instructions, and the orchestrator acts on them. The framework's defense is that imported state is tagged with its trust source; a constraint at the orchestrator's PAA can require that any state used to plan a high-stakes action be sourced from inside the trust boundary, or escalate.

\paragraph{Practical instantiation: the zero-trust agent gateway.} A canonical implementation pattern in enterprise settings is a \emph{zero-trust agent gateway}: a stateless component placed at the trust boundary that mediates every interaction between agents inside the deploying organization's trust scope and agents, tools, or content sources outside it. The gateway treats all inputs originating from outside the boundary as untrusted by default. State imported through the gateway is tagged with provenance metadata (source, authentication context, content classification), and constraints at the orchestrator's PAA require that any high-stakes action be planned over state whose tags satisfy a declared trust predicate. State that fails the predicate is either discarded, sanitized through a declared transformation, or escalated. The gateway is a runtime instantiation of SARC's enforcement surface extended across organizational boundaries: it brings the same constraint-class, predicate, and response semantics that govern internal agent dispatch to the cross-organizational interface, preserving SARC's internal enforcement assumptions rather than dissolving them at the perimeter. This pattern handles the practical case in which a deploying organization integrates third-party agents (vendor copilots, external API tools) without conceding the trust assumptions on which the SARC-conformant deployment depends.

\paragraph{Formalization: external state as a constraint subject.} The gateway pattern admits a clean formalization within the framework. External state crossing a trust boundary is treated as a constraint subject. Let $\tau(s_{\text{ext}})$ denote a trust predicate over imported state that incorporates provenance, authentication context, and content classification. For any high-stakes action $a$ that depends on $s_{\text{ext}}$, if $\tau(s_{\text{ext}}) = \text{false}$ then the corresponding constraint fires as $c \in C_e$ or $c \in C_h$ depending on the action's reversibility and domain risk. External-state usage is therefore governed by the same $\langle \texttt{src}, \texttt{class}, \texttt{pred}, \texttt{verif}, \texttt{resp} \rangle$ semantics as internal action constraints. The gateway is not a separate governance layer; it is the runtime location at which the trust predicate is evaluated and the standard SARC enforcement machinery is invoked. This collapses what is often treated as a perimeter security concern into the framework's existing constraint vocabulary, and ensures that crossing a trust boundary cannot bypass the enforcement surface that governs internal action.

\paragraph{Attribution dilution.} A workflow with deep recursion produces an audit trail in which every individual action is attributable but the relationship to the originating principal becomes opaque after $k$ hops. The trace is technically complete but governance-illegible. The framework's defense is the requirement that the trace is a tree, not a sequence, and that worker attribution tuples are preserved rather than summarized (Algorithm~\ref{alg:orch}, line 18). Attribution dilution is the multi-agent counterpart of the single-agent audit-by-reconstruction anti-pattern (\S\ref{sec:bolted}).

\subsection{Identity, Delegation, and Attribution}\label{sec:attribution}

The most consequential governance question in multi-agent systems is not optimization, composition, or conflict resolution; it is attribution. Who requested the action, which agent planned it, which agent executed it, which tool was invoked, under whose authority, and against which constraint set? An agent acting on behalf of a user is not the user, and the trace must distinguish them.

We require every entry in the runtime trace to carry an attribution tuple
\[
    \alpha = \langle \mathbf{P}, \texttt{planner}, \texttt{executor}, \texttt{tool}, \texttt{auth}, C_{\text{eval}} \rangle
\]
where $\mathbf{P} = [p_0, p_1, \ldots, p_k]$ is the principal-and-agent call chain (with $p_0$ the originating principal and $p_i$ for $i > 0$ the agent at depth $i$), \texttt{planner} is the agent that selected the action, \texttt{executor} is the agent that issued the tool invocation, \texttt{tool} is the specific tool called, \texttt{auth} is the intersected authority along $\mathbf{P}$ at the time of execution, and $C_{\text{eval}}$ is the set of constraints actually evaluated for this call with their outcomes.

Three properties follow. First, \emph{non-repudiation}: an action cannot be attributed to a principal who did not authorize the chain that produced it. Second, \emph{delegated authority is bounded}: a sub-agent inherits authority only to the extent of the chain's intersection at execution time; revocation at any layer truncates downstream authority. Third, \emph{constraint-set provenance}: any post-hoc audit can determine, for any action in the trace, which constraints were active and which evaluated, which is what makes Article 72 reporting tractable for multi-agent systems where a single user request fans out across multiple planners and executors.

The implication for the orchestrator-worker pattern is that the orchestrator's trace must include the worker's full attribution tuple, not merely a summary of the worker's output. Without that, a multi-agent system can satisfy the syntactic form of an audit log while obscuring the substantive question of who acted on whose behalf with what authority. A workflow that summarizes worker output into ``the agent decided $X$'' has lost the attribution information that distinguishes a delegated decision from an autonomous one, which under EU AI Act Art. 14 is precisely the distinction that determines whether human oversight has been effective.

\subsection{Conflict Resolution}

Cross-agent constraint conflicts (one agent's soft constraint to maximize throughput against another's hard constraint on freshness; one principal's authority to approve against another's authority to veto) are not resolved by the framework; they are surfaced. SARC's commitment is that conflicts cannot be silently resolved by an LLM at runtime; they must be either pre-resolved at specification time (precedence rules), or escalated. We treat conflict-resolution policy as part of the specification, declared at deployment, and reject the alternative of letting agents negotiate conflicts in natural language because the resolution is then unauditable.

For multi-principal conflicts (the all-of versus any-of question of \S\ref{sec:multi-auth}), the same rule applies: the composition is declared at specification time, and a runtime determination that a non-declared composition should apply triggers escalation rather than agent-mediated resolution.

\subsection{Constraint Composition Rules}\label{sec:multi-rules}

We close with a summary of the composition rules the preceding subsections have established. Let $\mathcal{A} = \{\Sigma_1, \ldots, \Sigma_n\}$ be the set of agent specifications in the workflow.

\begin{enumerate}[leftmargin=2em,topsep=0.3em,itemsep=0.3em]
\item \textbf{Inheritance ($\preceq$).} If $\Sigma_j$ is invoked by $\Sigma_i$, then $C_i \preceq C_j$: $\Sigma_j$ inherits all constraints of $\Sigma_i$ decidable in $\Sigma_j$'s state space. Non-decidable inherited constraints are evaluated at the deepest layer (closest to execution) at which the constraint remains decidable, or escalated (\S\ref{sec:multi-prop}).
\item \textbf{Disjunction for soft constraints ($\sqcup_s$).} Overlapping soft predicates compose by maximum cost, not sum, to avoid double-counting where a single regulatory source is referenced by multiple agents.
\item \textbf{Conjunction for hard constraints ($\sqcap_h$).} Hard constraints compose conjunctively: an action is admissible only if no agent's hard constraint blocks it. Dead-ends escalate.
\item \textbf{Authority intersection across the call chain.} A depth-$k$ action executes under $\bigcap_{i=0}^{k} \text{authority}(p_i)$; empty intersections fail-safe (Algorithm~\ref{alg:orch}, line 13).
\item \textbf{Multi-principal authority composition.} All-of by default; any-of must be declared with its qualifying-principal predicate (\S\ref{sec:multi-auth}).
\item \textbf{Precedence on escalations.} Source severity (regulatory > contractual > operational), with ties broken by lexicographic source ordering declared at deployment time.
\end{enumerate}

These rules are necessary but not in themselves sufficient. The multi-agent setting raises two open questions the framework does not yet resolve. First, recursive workflows in which the depth $k$ is itself state-dependent (an agent decides at runtime whether to invoke a sub-agent) admit attack surfaces (an adversary controlling the depth-decision policy can stretch the call chain to dilute attribution) that we have not formally characterized. Second, market and auction topologies in which counterparty agents are outside the deploying organization's trust boundary require constraint-on-protocol semantics that this paper has only sketched. We treat both as future work that the framework as developed here is positioned to address.

\section{Empirical Evaluation: SARC vs.\ Practitioner Baselines}\label{sec:sim}

This section reports a reproducible synthetic evaluation comparing SARC against four practitioner baselines (post-hoc audit, output filtering, workflow rules, policy-as-code-only) on a procurement task. The evaluation is synthetic, not a deployment study, and we are explicit about its scope: it isolates the structural difference between governance regimes on a controlled task, characterizes the latency and escalation costs SARC pays, and provides controlled empirical support for the residual-violation scaling argument. It does not establish that SARC will produce the same numbers in real procurement, customer-service, or clinical settings; deployment evaluation through staged rollout (shadow execution, partial-traffic dual-running, full cutover) is left to the follow-up work scoped in \S\ref{sec:future-impl}. What it shows, with reported confidence intervals over 50 random seeds, is that the framework's enforcement-timing claim and residual-violation scaling claim are supported in a controlled multi-baseline measurement.

\paragraph{Why procurement, and why this matters beyond procurement.} The procurement example is stylized, but the constraint pattern it exhibits is common in production agentic systems. Customer-support agents face refund thresholds and escalation windows. Internal copilots face document-access and data-leakage constraints. Procurement agents face supplier, sanctions, and approval constraints. HR agents face protected-class and employment-decision constraints. Clinical decision-support tools face scope-of-use and uncertainty-threshold constraints. In each case, the central governance question is not whether the agent can generate an acceptable response, but whether it can be prevented from executing an inadmissible action and whether the resulting trace can substantiate what happened. The procurement instance is a clean controlled case for evaluation; the structural lesson is expected to transfer to other domains but requires further empirical testing.

\subsection{Setup}

A synthetic procurement environment with $H = 1{,}000$ orders per episode. Order amounts are sampled from a log-normal distribution ($\mu_{\ln} = 8.5$, $\sigma_{\ln} = 1.2$; median $\approx$ \texteuro{}4{,}900; right tail past \texteuro{}50{,}000). First-time-supplier flag is drawn independently with probability $0.135$ per order. Three constraints apply: a hard constraint $c_h$ (orders above \texteuro{}50{,}000 require approval), a soft constraint $c_s$ (rolling 24-hour spend at or above \texteuro{}475{,}000 should trigger throttling), and an escalation constraint $c_e$ (first-time suppliers require review). The agent's reward is total spend successfully placed, a deliberately misaligned proxy that creates pressure to violate constraints.

\subsection{Baselines}\label{sec:sim-baselines}

We compare five regimes across 50 random seeds.

\begin{itemize}[leftmargin=2em,topsep=0.3em,itemsep=0.2em]
\item \textbf{Post-hoc audit.} Constraints checked only after execution. Documents violations; does not prevent them.
\item \textbf{Output filter.} A lightweight text-level filter that catches some visible high-risk behavior but cannot reliably see side-effectful tool calls. We model partial coverage of $c_h$ at 25\% post-action.
\item \textbf{Workflow rules.} Static orchestration rules with an if/else gate on $c_h$ before tool dispatch. Catches high-value purchase orders but does not provide full constraint-class coverage of $c_s$ and $c_e$ semantics.
\item \textbf{Policy-as-code-only.} A generic policy decision engine evaluating $c_h$ at the policy layer and $c_e$ at PAG. The baseline does not declare a PAA throttling response for $c_s$, and does not emit a SARC-conformant trace structure for audit checking.
\item \textbf{SARC.} $c_h$ at PAG (block-or-escalate), $c_e$ at PAG (suspend-route-default-deny), $c_s$ at PAA (throttle-log) with a declared throttling response that backs off subsequent rolling-spend exposure after a completed action crosses the soft window. Structured trace emission and audit checking apply.
\end{itemize}

Per-step latency is modeled as a fixed cost added by each regime's enforcement layer. We do not claim these latencies match a specific real implementation; they capture the expected ordering (post-hoc audit cheapest, SARC most expensive due to multiple class-typed evaluation points) and let us report the safety/latency trade-off as a measured curve rather than as a stated assumption. Operator capacity is set to $c=2$ with mean service time 6 minutes, yielding $\rho \approx 0.42$ at the observed escalation arrival rate (well below saturation).

\subsection{Results}\label{sec:sim-results}

Across 50 seeds (mean $\pm$ 95\% confidence interval), the five regimes exhibit the structurally expected pattern.

\begin{table}[!htbp]
\centering\footnotesize
\begin{tabular}{@{}lrrrrr@{}}
\toprule
\textbf{Regime} & \textbf{Hard exec.} & \textbf{Soft overages} & \textbf{Supp.\ no review} & \textbf{Escalations} & \textbf{Latency / step} \\
\midrule
Post-hoc audit       & $26.8 \pm 1.4$  & $949.8 \pm 2.9$ & $132.3 \pm 2.4$ & $0.0 \pm 0.0$    &  0 ms \\
Output filter        & $19.4 \pm 1.1$  & $947.4 \pm 3.7$ & $132.7 \pm 2.3$ & $0.0 \pm 0.0$    &  7 ms \\
Workflow rules       & $0.0 \pm 0.0$   & $935.6 \pm 2.3$ & $129.0 \pm 2.6$ & $0.0 \pm 0.0$    & 12 ms \\
Policy-as-code-only  & $0.0 \pm 0.0$   & $937.1 \pm 2.1$ & $0.0 \pm 0.0$   & $133.8 \pm 2.5$  & 15 ms \\
\textbf{SARC}        & $0.0 \pm 0.0$   & $\mathbf{98.8 \pm 1.0}$ & $0.0 \pm 0.0$ & $160.2 \pm 3.3$ & 21 ms \\
\bottomrule
\end{tabular}
\caption{Synthetic procurement task, $H = 1{,}000$ orders per episode, $N = 50$ seeds, mean $\pm$ 95\% CI. Hard exec.\ = hard-constraint violations executed; Soft overages = rolling-spend window exceedances; Supp.\ no review = first-time suppliers cleared without review; Escalations = constraint events routed to ER; Latency / step = per-step enforcement overhead.}
\label{tab:sim}
\end{table}

\paragraph{Interpretation.} Across 50 seeds, SARC and policy-as-code-only both block executed high-value hard violations under exact predicates. SARC does not claim that no other architecture can block a known hard rule; the difference is in class-complete governance. Workflow rules and output filtering catch $c_h$ partially or fully but leave the supplier-review path uncovered. Policy-as-code-only adds escalation for $c_e$ but does not declare a PAA throttling response for the rolling-spend soft window. SARC declares all three responses at class-compatible enforcement points and emits a trace structure the audit checker of \S\ref{sec:audit-checker} verifies for every action.

\paragraph{Reading the soft-window result carefully.} SARC reduces soft-window overages from $\sim$937 (policy-as-code-only) to $98.8 \pm 1.0$, a $89.5\%$ reduction at $95\%$ confidence. This number requires careful framing. The reduction is a consequence of the \emph{declared PAA throttling response} that backs off subsequent rolling-spend exposure after a completed action crosses the soft window. It is not a magic property of the SARC label. A policy-as-code system with the same PAA-positioned throttling response would be expected to show a similar reduction. The architectural claim SARC makes is more specific: the framework requires this response to be declared, class-typed, and placed at the runtime point at which the predicate is most meaningful, namely over completed-action state at PAA rather than over proposed-action state at PAG. The 89.5\% figure is therefore evidence that placement-discipline-and-typing has measurable consequences on a controlled task, not evidence that SARC has a built-in throttling capability that other architectures lack. The result should accordingly be interpreted as evidence that class-typed placement makes response semantics explicit and measurable, not as evidence that SARC intrinsically outperforms any policy-as-code implementation.

\paragraph{Costs SARC pays.} The framework forces these costs to be visible. Per-step latency rises from 0~ms (post-hoc) to 21~ms (SARC). Escalation load rises from zero (post-hoc, output filter, workflow rules; no escalation surface declared) to $133.8 \pm 2.5$ for policy-as-code-only and $160.2 \pm 3.3$ for SARC. These are the trade-offs the architecture is meant to surface: runtime governance is not free, but its costs are measurable and admit explicit operational planning (operator staffing, latency budget allocation), rather than emerging as production surprises.

\paragraph{Audit checker exercised on every SARC trace.} The audit checker of \S\ref{sec:audit-checker} runs on the runtime trace produced by every SARC seed. Across the 50 seeds, the checker reports zero structural discrepancies on the SARC traces: every applicable constraint is evaluated at a class-compatible point, every fired constraint produces the declared response, and every action carries a non-empty attribution chain. This is the empirical instantiation of the decidable-audit property of \S\ref{sec:audit-checker}: the checker can answer ``did the system honor its specification?'' for finite traces in bounded time, given the specification and the trace alone.

\paragraph{Artifact availability.} The audit checker, benchmark harness, per-seed CSV outputs, and summary JSON are provided as supplementary artifacts; the evaluation tables in this section are computed directly from these files. The artifact bundle is available at \url{https://github.com/besanson/sarc-governance}.

\subsection{Sensitivity Analysis}\label{sec:sensitivity}

The headline numbers in Table~\ref{tab:sim} are point estimates. To establish that the structural claim (that runtime enforcement is qualitatively different from post-hoc audit) is not an artefact of the chosen parameters, we vary three: the underlying violation probability $\epsilon$, the escalation arrival rate $\lambda_e$, and operator capacity $c$.

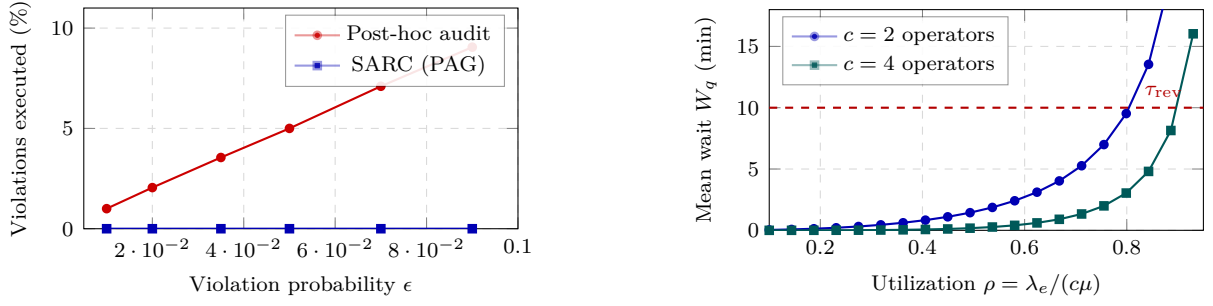
\begin{figure}[h]
\centering
\begin{tikzpicture}
\begin{axis}[
    width=0.46\textwidth, height=4.5cm,
    xlabel={Violation probability $\epsilon$},
    ylabel={Violations executed (\%)},
    xmin=0.005, xmax=0.10,
    ymin=0, ymax=11,
    legend style={font=\scriptsize, at={(0.97,0.97)}, anchor=north east, fill=white, fill opacity=0.9, draw opacity=0.4, text opacity=1},
    tick label style={font=\scriptsize},
    label style={font=\scriptsize},
    grid=major, grid style={dashed,gray!30}
]
\addplot[color=red!80!black, thick, mark=*, mark size=1.4pt] coordinates {
 (0.01, 1.0) (0.02, 2.05) (0.035, 3.55) (0.05, 5.0) (0.07, 7.1) (0.09, 9.05)
};
\addlegendentry{Post-hoc audit}
\addplot[color=blue!70!black, thick, mark=square*, mark size=1.4pt] coordinates {
 (0.01, 0) (0.02, 0) (0.035, 0) (0.05, 0) (0.07, 0) (0.09, 0)
};
\addlegendentry{SARC (PAG)}
\end{axis}
\end{tikzpicture}\hfill
\begin{tikzpicture}
\begin{axis}[
    width=0.46\textwidth, height=4.5cm,
    xlabel={Utilization $\rho = \lambda_e/(c\mu)$},
    ylabel={Mean wait $W_q$ (min)},
    xmin=0.1, xmax=0.95,
    ymin=0, ymax=18,
    legend style={font=\scriptsize, at={(0.03,0.97)}, anchor=north west, fill=white, fill opacity=0.9, draw opacity=0.4, text opacity=1},
    tick label style={font=\scriptsize},
    label style={font=\scriptsize},
    grid=major, grid style={dashed,gray!30}
]
\addplot[color=blue!70!black, thick, mark=*, mark size=1.4pt, samples=20, domain=0.1:0.93] {6 * x*x / (1-x) / 2};
\addlegendentry{$c=2$ operators}
\addplot[color=teal!70!black, thick, mark=square*, mark size=1.4pt, samples=20, domain=0.1:0.93] {6 * x*x*x*x / (1-x) / 4};
\addlegendentry{$c=4$ operators}
\addplot[color=red!70!black, dashed, thick, samples=2, domain=0.1:0.95] {10};
\node[font=\scriptsize, color=red!70!black, anchor=south east] at (axis cs:0.93,10) {$\tau_{\text{rev}}$};
\end{axis}
\end{tikzpicture}
\caption{(Left) Hard-violation rate vs.\ underlying violation probability $\epsilon$. The post-hoc-audit curve scales linearly with $\epsilon$; the SARC curve is flat at zero because PAG blocks before dispatch. Under the modeled assumptions, the structural distinction is preserved as $\epsilon$ varies. (Right) Escalation queue mean wait $W_q$ vs.\ utilization $\rho$, M/M/c. As $\rho \to 1$, $W_q$ diverges super-linearly. The reversibility window $\tau_{\text{rev}}$ (dashed) defines the operationally admissible region. The architect's task is to keep $(\lambda_e, c, \mu)$ in the region where $W_q < \tau_{\text{rev}}$; once the system leaves that region, escalation is structurally too slow regardless of operator effort.}
\label{fig:sens-eps-rho}
\end{figure}

\begin{figure}[h]
\centering
\begin{tikzpicture}
\begin{axis}[
    width=0.6\textwidth, height=5cm,
    xlabel={Per-step latency overhead (ms)},
    ylabel={Hard-violation rate (\%)},
    xmin=0, xmax=120,
    ymin=-0.3, ymax=6,
    legend style={font=\scriptsize, at={(0.97,0.97)}, anchor=north east, fill=white, fill opacity=0.9, draw opacity=0.4, text opacity=1},
    tick label style={font=\scriptsize},
    label style={font=\scriptsize},
    grid=major, grid style={dashed,gray!30}
]
\addplot[only marks, mark=square*, mark size=2.2pt, color=red!80!black] coordinates { (0, 4.7) };
\node[font=\tiny, anchor=west] at (axis cs:2,4.7) {Audit only};
\addplot[only marks, mark=triangle*, mark size=2.5pt, color=orange!80!black] coordinates { (8, 2.1) };
\node[font=\tiny, anchor=west] at (axis cs:10,2.1) {Output filter};
\addplot[only marks, mark=diamond*, mark size=2.5pt, color=teal!70!black] coordinates { (15, 0.8) };
\node[font=\tiny, anchor=west] at (axis cs:17,0.8) {Workflow rules};
\addplot[only marks, mark=*, mark size=2.5pt, color=blue!70!black] coordinates { (21, 0) };
\node[font=\tiny, anchor=west] at (axis cs:23,0) {SARC (PAG)};
\addplot[only marks, mark=*, mark size=2.5pt, color=blue!50!black] coordinates { (45, 0) };
\node[font=\tiny, anchor=west] at (axis cs:47,0) {SARC + retrieval check};
\addplot[only marks, mark=*, mark size=2.5pt, color=blue!30!black] coordinates { (110, 0) };
\node[font=\tiny, anchor=west] at (axis cs:91,0.4) {SARC w/ LLM-judge check};
\end{axis}
\end{tikzpicture}
\caption{Latency-versus-safety trade-off across enforcement strategies. Post-hoc audit incurs zero per-step latency but pays the full violation rate. Output filtering and workflow rules cut violations partially at modest latency cost. SARC at PAG eliminates hard violations at $\sim$21~ms (deterministic predicate over typed signature). SARC with retrieval-augmented checks costs more but stays in O($\ell_{\text{tool}}$). SARC implementations that route checks through a separate LLM call are flagged for the architectural reason discussed in \S\ref{sec:latency}: they cross the O($\ell_{\text{model}}$) boundary and inherit the nondeterminism the check is meant to bound.}
\label{fig:sens-latency}
\end{figure}
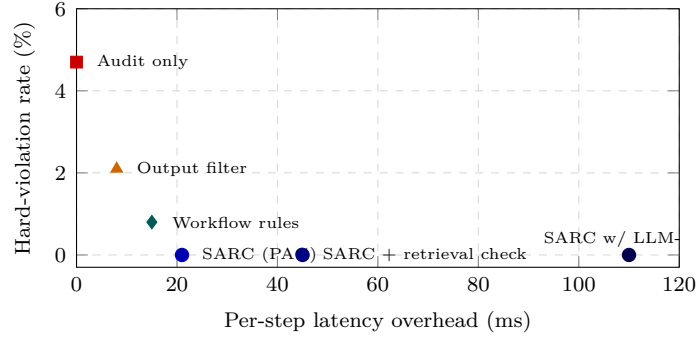

\paragraph{What the sensitivity analysis shows.} Three structural properties of the framework are robust across the parameter sweeps. First, the SARC-vs-audit gap on hard violations is qualitative (zero versus linear in $\epsilon$), not quantitative; tuning $\epsilon$ does not close it. Second, the escalation queue has a defined operationally admissible region governed by $W_q < \tau_{\text{rev}}$; outside that region, no amount of constraint specification rescues the deployment, and the architect must either staff up, tighten the predicate, or accept that escalation is not a viable response class for this action. Third, the latency cost of SARC enforcement is not a single number but a function of the implementation choice; deterministic predicates over typed signatures are cheap, retrieval-augmented checks are intermediate, and LLM-as-judge checks cross a structural threshold that the framework explicitly warns against. The trade-off curve makes the architectural choice quantifiable.

\paragraph{Generated versus analytical artifacts.} For clarity about which results come from which source, Figures~\ref{fig:sens-eps-rho} and~\ref{fig:sens-latency} are analytical sensitivity illustrations parameterized from the benchmark assumptions: the post-hoc-audit curve is the linear scaling implied by the violation-opportunity model, the SARC curve is the structural zero implied by PAG admissibility under exact predicates, the queueing curve is the M/M/c Erlang-C wait formula at the declared service rate, and the latency-versus-safety scatter is the regime-by-regime ordering implied by the same per-step latency parameters used in the benchmark. Tables~\ref{tab:sim} and~\ref{tab:residual}, by contrast, report per-seed simulation results generated directly by the released evaluation harness over $N=50$ seeds with reported 95\% confidence intervals.

\paragraph{Caveat on the zero-violation curve.} The flat-zero SARC curve in Figure~\ref{fig:sens-eps-rho} (left) assumes exact predicate evaluation, active enforcement at every step, and the absence of tool-layer or policy-layer bypass paths. If predicate error, enforcement-process failure, or bypass is introduced, the SARC curve becomes a function of those failure probabilities rather than of the underlying violation opportunity rate $\epsilon$ alone. The structural claim of the paper is that the dependence on $\epsilon$ is removed by SARC, not that residual violations are impossible; residual violations under SARC arise from implementation defects in $C$, not from the underlying decision problem.

\paragraph{Threats to validity.} The evaluation has three threats to validity that should be named explicitly. First, the environment is synthetic and intentionally stylized; results from a synthetic procurement task do not generalize to real procurement, customer-service, or clinical settings without further empirical work. Second, the per-step latency figures are model parameters of the regimes rather than measurements of any specific implementation; they capture the expected ordering, not the absolute latency a specific deployment will experience. Third, the agent's reward function is deliberately misaligned to expose the gap between runtime enforcement and post-hoc audit; under a well-aligned reward the gap may narrow even without SARC, though the structural argument of \S\ref{sec:counter} suggests it does not close. The evaluation supports a structural-and-empirical claim about enforcement timing and class-typed response placement; it does not establish absolute deployment-grade performance numbers.

\subsection{Residual Violations Under Imperfect Enforcement}\label{sec:residual}

The zero-violation result under SARC should be read as an idealized enforcement result, not as a claim that residual violations are impossible. Under SARC, residual hard-constraint violations are not driven by the opportunity rate of violations in the environment, but by implementation error in the enforcement stack. Let $\epsilon_{\text{pred}}$ denote predicate error (false negatives in the predicate evaluation itself) and $\epsilon_{\text{exec}}$ denote enforcement-process failure, including tool-layer bypass, policy misconfiguration, or monitor unavailability. In the idealized case used in the headline results, $\epsilon_{\text{pred}} = \epsilon_{\text{exec}} = 0$, so hard violations are blocked before dispatch. When these error terms are non-zero, residual violations scale with $\epsilon_{\text{pred}}$ and $\epsilon_{\text{exec}}$ rather than with the raw environmental opportunity rate. We measure this directly.

\begin{table}[!htbp]
\centering\footnotesize
\begin{tabular}{@{}rrrr@{}}
\toprule
\textbf{Predicate FN rate $\epsilon_{\text{pred}}$} & \textbf{Enforcement failure rate $\epsilon_{\text{exec}}$} & \textbf{Hard violations executed} & \textbf{(mean $\pm$ 95\% CI)} \\
\midrule
$0.00$ & $0.00$ & $0.00$  & $\pm$ $0.00$ \\
$0.00$ & $0.01$ & $0.26$  & $\pm$ $0.15$ \\
$0.00$ & $0.05$ & $1.20$  & $\pm$ $0.31$ \\
$0.01$ & $0.00$ & $0.20$  & $\pm$ $0.14$ \\
$0.01$ & $0.01$ & $0.46$  & $\pm$ $0.20$ \\
$0.01$ & $0.05$ & $1.38$  & $\pm$ $0.36$ \\
$0.05$ & $0.00$ & $1.32$  & $\pm$ $0.35$ \\
$0.05$ & $0.01$ & $1.58$  & $\pm$ $0.37$ \\
$0.05$ & $0.05$ & $2.50$  & $\pm$ $0.46$ \\
$0.10$ & $0.00$ & $2.52$  & $\pm$ $0.50$ \\
$0.10$ & $0.01$ & $2.78$  & $\pm$ $0.51$ \\
$0.10$ & $0.05$ & $3.66$  & $\pm$ $0.59$ \\
\bottomrule
\end{tabular}
\caption{Predicate-noise / enforcement-failure sweep under SARC, $H = 1{,}000$ orders per episode, $N = 50$ seeds per cell, mean $\pm$ 95\% CI. Under exact predicates and perfect enforcement (top row), SARC executes zero hard violations. Residual violations grow approximately linearly in the enforcement-stack error budget $\epsilon_{\text{pred}} + \epsilon_{\text{exec}}$ rather than in the environmental opportunity rate, which under post-hoc audit yields $\sim$26.8 hard executions per episode (Table~\ref{tab:sim}). The structural property the framework targets is decoupling residual violations from environmental pressure.}
\label{tab:residual}
\end{table}

The table substantiates the framework's structural claim empirically. At $\epsilon_{\text{pred}} = \epsilon_{\text{exec}} = 0$, SARC executes zero hard violations against a baseline post-hoc-audit rate of $26.8$ per episode (Table~\ref{tab:sim}). At $\epsilon_{\text{pred}} = 0.10$, $\epsilon_{\text{exec}} = 0.05$, residual violations rise to $3.66$ per episode, an order of magnitude below the post-hoc-audit baseline despite both noise sources being non-trivial. The implication for engineering practice is that the cost of reducing residual violations under SARC is the cost of reducing $\epsilon_{\text{pred}}$ and $\epsilon_{\text{exec}}$, both of which are addressable through standard software-assurance practices: predicate testing, redundant enforcement points, monitor health checks, and structured incident response when monitors fail. Under post-hoc audit, the cost of reducing residual violations is the cost of reducing the opportunity rate $\epsilon$ itself, which is rarely addressable without reducing the system's utility.

\paragraph{What this evaluation does not show.} It does not show that SARC outperforms safe-RL training. It does not show generalization to real customer-service, clinical, or HR settings without further empirical work. It does not include an adversarial evaluation against indirect prompt injection: the AgentDojo-style suite scoped in \S\ref{sec:future-impl} would test the framework's defensive value against the trust-boundary attacks of \S\ref{sec:multi-failure}, which the present evaluation does not stress. In enterprise deployments, additional validation is likely to proceed through staged rollout strategies (shadow execution alongside an existing system, then partial-traffic dual-running, then full cutover); such methodologies are outside the scope of this work but are the natural next step for organizations operationalizing the framework.

\section{Failure Modes: Bolted-On vs Built-In Constraints}\label{sec:bolted}

\paragraph{The wrapper anti-pattern.} Constraint checks placed in a wrapper around an LLM call, leaving the orchestration layer untouched. Honors $C_h$ at most weakly: the action space is the tool registry, not the LLM output, and a model instructed to act via tools can route around an output filter.

\paragraph{The dashboard anti-pattern.} Operator presented with post-hoc alerts. Satisfies the documentation form of Art.~14 without its substance. The vigilance literature \cite{parasuraman1997vigilance,anaby2024alertfatigue} shows operators presented with high-rate post-hoc alerts either over-acknowledge (rubber-stamping) or under-acknowledge (alert fatigue); neither produces effective oversight. SARC requires PAG-time predictive escalation for $C_e$.

\paragraph{The reward-shaping anti-pattern.} Constraints folded into $R$ as large negative shaping terms. Proposition~\ref{prop:primacy} and Example~\ref{ex:counter} address this directly. The pattern is unstable for the structural reason given in the counterexample: low-probability high-reward events expose the finiteness of the penalty. The AI-safety literature has documented the operational consequences extensively under the rubric of specification gaming and reward hacking \cite{amodei2016concrete,leike2017gridworlds,krakovna2020specgaming,skalse2022rewardhacking,hendrycks2021unsolved}: agents reliably find policies that optimize the proxy reward without satisfying the underlying intent. SARC's response is architectural rather than algorithmic (constraints belong outside the reward) but the failure mode the architectural choice avoids is precisely the one this literature characterizes.

\paragraph{The audit-by-reconstruction anti-pattern.} Compliance produced by inspecting logs after the fact. Structurally fragile: changes to the regulation, the system, or the logging schema invalidate the mapping silently. SARC inverts the dependency: audit trail generated from the specification. The auditing literature has long identified the same dependency for human-facing AI systems, model cards \cite{mitchell2019modelcards}, datasheets \cite{gebru2021datasheets}, and end-to-end algorithmic auditing frameworks \cite{raji2020closing,raji2022outsider} are all efforts to attach documentation to models in a way that survives system change. SARC extends the same logic into the agent runtime: documentation is not appended to a deployed system; it is emitted by it.

The common pattern: each failure mode implements the constraint at a layer or stage that does not honor its class semantics. The discipline SARC imposes is to refuse this collapse.

\section{Economic Implications of Runtime Governance}\label{sec:economic}

The operational trade-offs of \S\ref{sec:ops} have economic content that the foregoing has treated implicitly. We make it explicit here, briefly, because the architectural choices SARC forces are also capital allocation decisions under risk.

\subsection{Three Cost Channels and the Operating Point}

Every constraint $c \in C$ generates costs along three channels: false-positive cost $\kappa_{\text{FP}}$ (lost throughput when an admissible action is blocked), false-negative cost $\kappa_{\text{FN}}$ (regulatory exposure, operational consequence, reputational impact when an inadmissible action permits a violation), and escalation cost $\kappa_{\text{ER}}$ (operator time and added wait latency, scaling super-linearly as queue utilization $\rho \to 1$ per \S\ref{sec:queue}). For any predicate, the architect chooses an operating point $\theta$ minimizing
\[
    \mathbb{E}[\text{cost}(\theta)] = \kappa_{\text{FP}} \cdot p_{\text{FP}}(\theta) + \kappa_{\text{FN}} \cdot p_{\text{FN}}(\theta) + \kappa_{\text{ER}} \cdot p_{\text{esc}}(\theta).
\]
The optimal $\theta^*$ depends on the firm's risk tolerance (rarely an engineering parameter and frequently absent from technical specifications), on operator capacity (which couples constraint calibration to staffing decisions), and on action reversibility (shorter reversibility windows justify more conservative operating points because the false-negative cost is realized faster). Asymmetric loss settings, where $\kappa_{\text{FN}} \gg \kappa_{\text{FP}}$, drive $\theta^*$ toward conservative enforcement; a small numerical illustration of this asymmetry is given in Appendix~\ref{app:economic}.

\paragraph{Constraint calibration is a capital allocation decision under risk.} The expression for $\theta^*$ is not a tuning problem. It is the architect's declaration of how the firm allocates capital among false-positive cost, false-negative cost, and operator cost. A SARC specification that does not name the operating point of each predicate is under-specified in a fiduciary sense, not only a technical one. The framework operationalizes this through invariants I2 (every constraint declares an operating point) and I4 (every escalation declares a reversibility window with bounded queueing latency relative to it); the economic argument is therefore a second reading of the same invariants that govern conformance.

\subsection{Implications for Procurement and Build Decisions}

Two implications follow. First, vendor evaluation must include constraint specification quality, not only model performance. A vendor offering an agentic system whose hard constraints are hosted at the prompt layer is offering a system whose false-negative tail is uncharacterized; the difference between $C_h$ at the tool or policy layer and $C_h$ at the prompt layer is the difference between bounded and unbounded tail risk on the deployment. Second, internal build decisions must budget for the operator capacity the constraint specification implies. The escalation rate observed in \S\ref{sec:sim} (160 escalations per 1{,}000-order episode under SARC) at a service time of six minutes per case implies operator staffing as a first-order term in the deployment's cost, not an afterthought.

\section{Conclusion and Open Directions}\label{sec:concl}

SARC is a small framework: one tuple, four components, four runtime enforcement points, one compilation path. The contribution is in what the framework refuses, the collapse of constraints into rewards, the deferral of hard checks to post-hoc audit, the treatment of human oversight as a dashboard, the production of compliance documentation as an artifact disjoint from the system that operates. In each case, the refusal compels work upstream: at specification time, in the runtime architecture, at the compilation interface between obligation and check.

The framework is intentionally compatible with the optimization machinery of safe-RL. A SARC specification is meant to compile into a CMDP where the underlying problem admits that treatment; the safe-RL machinery solves the result. SARC delivers that machinery a problem that is fully posed, in the dimensions on which sustainable governable deployment depends.

\subsection{The Thesis Line}

SARC does not make systems safe. It makes safety expressible, enforceable, and auditable at runtime, which is the condition under which safety can be engineered. The distinction is operational: systems fail, but under SARC they fail as violations of declared constraints rather than as unobservable anomalies. A system that fails as a declared violation can be corrected; a system that fails as an anomaly can only be apologized for.

\paragraph{What SARC separates.} The framework separates three concerns that are otherwise entangled in agentic systems: \emph{optimization} (R), \emph{capability} (A), and \emph{admissibility} (C). Optimization is what the system tries to do well; capability is what it can do; admissibility is what it is permitted to do. Conflating these is the structural error that leads systems to fail as anomalies. Separating them is what allows the same system to be optimized, extended, and governed by different stakeholders along different timescales without the three concerns silently coupling at runtime.

\paragraph{Architectural inevitability in Regime B.} A second observation follows from Proposition~\ref{prop:primacy} and the failure-mode taxonomy of \S\ref{sec:bolted}. In Regime B (the deployment regime characterized by regulatory exposure, irreversible actions, or asymmetric loss), the question becomes less whether to implement SARC and more where in the system equivalent enforcement will emerge. Systems that do not adopt an explicit SARC structure tend to reconstruct equivalent functions through ad hoc checks, periodic audits, on-call escalation rotations, and logging conventions reconstructed into compliance reports after the fact. The functions are often present; what is absent is the architectural discipline that makes them coherent, machine-checkable, and durable across staff turnover and regulatory change. This paper argues that, in Regime B, organizations tend to face a choice between explicit runtime-governance architecture and ad hoc accretion of equivalent controls, and that the latter pattern is what we observe in current practice.

\subsection{Limits of the Framework}\label{sec:limits}

To preserve the precision of the thesis line, we name what SARC does not do. The framework does not solve legal interpretation: it assumes that obligations have been translated from statutory text into operationally evaluable predicates. It does not solve predicate correctness: a predicate that miscaptures the obligation produces a system that complies with the wrong rule. It does not solve model reliability: the LLM at the planning layer can still fail, drift, or hallucinate, and SARC bounds the consequences rather than prevents the underlying error. It does not solve organizational accountability: an unowned constraint that nobody maintains will decay regardless of its architectural placement. It does not replace data governance, model evaluation, conformity assessment, or institutional oversight; it provides the runtime enforcement surface on which those upstream activities rest.

The framework's value is precisely its narrowness. SARC governs the runtime enforcement surface and emits the audit artifacts that make the upstream work checkable. It is the substrate; it is not the whole governance program. A deploying organization that adopts SARC and neglects predicate maintenance, ownership assignment, or upstream legal interpretation will end up with a system that fails as declared constraint violations against a specification that no longer reflects the obligation. That is still better than failing as anomalies, but it is not the system SARC was designed to enable.

\subsection{Adoption Ladder}\label{sec:adoption}

The framework as developed in this paper is comprehensive. We do not expect (or recommend) that an organization adopt it in full from a standing start. We propose a four-level adoption ladder that converts the framework from a single normative target into an incremental implementation path.

\begin{description}[leftmargin=2.6em,topsep=0.3em,itemsep=0.4em]
\item[\textbf{Level 1, Minimal.}] Single-agent deployment. Pre-Action Gate only. Hard constraints ($C_h$) only, hosted at the tool or policy layer. Structured trace emission for every action. This level is sufficient to satisfy invariants I1, I2 (for $C_h$ alone), I3, I6, I7, and I8 of \S\ref{sec:invariants}.
\item[\textbf{Level 2, Intermediate.}] Add Post-Action Auditor. Add soft constraints ($C_s$) with calibrated operating points. Add basic telemetry conforming to Art.~72 schema. The system now characterizes its own behavior over time and can support drift detection.
\item[\textbf{Level 3, Advanced.}] Add Action-Time Monitor. Add escalation constraints ($C_e$) with declared reversibility windows and an Escalation Router with declared queueing capacity. The system now satisfies all four enforcement points of the reference architecture (\S\ref{sec:arch}) and can support an Article 14 effective-human-oversight argument under the framework's interpretation.
\item[\textbf{Level 4, Full SARC.}] Multi-agent. Constraint propagation under recursion. Authority composition across principals. Attribution-preserving trace trees. Cross-agent failure-mode defenses. The system is now governable at the workflow level, not only at the individual agent level, and is positioned for the recursive and multi-principal cases that characterize real enterprise deployment.
\end{description}

The adoption ladder is not a maturity model in the sense of CMM or its successors. It is a sequencing recommendation: each level is an internally coherent partial implementation, and an organization at Level $n$ has working governance over the surface its level covers. Stepping up to Level $n+1$ is a deliberate decision that adds capability without invalidating what is already in place.

\subsection{Implementation Status and Remaining Work}\label{sec:future-impl}

This paper specifies the architecture, formalizes the audit property, and reports a synthetic evaluation against four practitioner baselines with reported confidence intervals. We name here what is included in the present artifact and what remains for future work.

\paragraph{Included in this paper.} A prototype audit checker (Python, included as supplementary material at \url{https://github.com/besanson/sarc-governance}) that verifies $T \models \Sigma$ over finite specifications and traces per Definition~\ref{def:stc}. A reproducible synthetic evaluation harness comparing SARC against post-hoc audit, output filtering, workflow rules, and policy-as-code-only baselines on a procurement task, run over $N=50$ random seeds with reported 95\% confidence intervals (\S\ref{sec:sim}). A predicate-noise / enforcement-failure sweep over twelve $(\epsilon_{\text{pred}}, \epsilon_{\text{exec}})$ cells with $N=50$ seeds per cell, supporting the residual-violation scaling argument under controlled assumptions (\S\ref{sec:residual}). The audit checker is exercised on the runtime trace produced by every SARC seed and reports zero structural discrepancies.

\paragraph{Not included; the natural next steps.} An integration of the prototype with a contemporary agent runtime such as LangGraph or AutoGen, exposing PAG, ATM, PAA, and ER as hooks against a typed tool registry. An adversarial evaluation drawing on AgentDojo \cite{debenedetti2024agentdojo} and follow-on adaptive-attack benchmarks against indirect prompt injection in tool-using agents, scoring each baseline on attack-success rate, false-positive rate on legitimate tasks, escalation load, and latency under attack. Evaluation on at least one additional task domain (a benefit-credit copilot, a refund-processing agent, or a clinical decision-support task) to test how the structural lessons here transfer beyond the procurement instance. A real-deployment shadow rollout against an existing production agent in a regulated workflow, characterizing the gap between synthetic and operational predicate noise.

The present paper supports the framework's structural and empirical claims on a controlled task with reported confidence intervals, an implemented and exercised audit checker, and a measured residual-violation sweep. The remaining work is empirical extension into additional domains and adversarial settings; it does not require revisiting the framework or its formal properties.

\subsection{Open Directions}

Three directions remain open. First, formal compilation from a SARC specification to a CMDP, including completeness conditions and information-loss profile. Second, broader empirical validation on real deployments, the simulation in \S\ref{sec:sim} is a starting point, not an endpoint. Third, the multi-agent treatment in \S\ref{sec:multi} addresses the typical enterprise topologies but leaves two open subproblems flagged at the end of \S\ref{sec:multi-rules}: state-dependent recursion depth as an attack surface for attribution dilution, and constraint-on-protocol semantics for market and auction topologies in which counterparty agents are outside the deploying organization's trust boundary. Both deserve dedicated treatment.

The vocabulary a field uses to describe its objects determines the failures the field can see. Under $\langle S, A, R \rangle$, agentic failures in regulated deployment are visible only as anomalies. Under $\langle S, A, R, C \rangle$, with $C$ enforced in the runtime, those failures become violations of declared constraints with named source, named predicate, named response. SARC removes a blind spot, and provides the architectural slot in which governance can live.

\appendix

\section{Sample SARC Specification (YAML)}\label{app:yaml}

To make the framework concrete, we present a sample SARC specification for a procurement-approval agent in YAML. The specification is intended as a reference format, not as a normative serialization standard. The fields correspond directly to the constructs of \S\ref{sec:def}--\S\ref{sec:arch}: each constraint declares its source, class, predicate, verification point, response, escalation routing, default-on-timeout behavior, and the trace fields it emits.

\begin{small}
\begin{verbatim}
spec_version: "sarc-0.1"
agent: procurement-approver
deployment: "enterprise procurement workflow with regulatory, contractual, and operational constraints; illustrative example, not a claim of EU AI Act Annex III classification"

state:
  modalities: [purchase_order, supplier_record, budget_state]
  retrieval:
    - source: erp.purchase_orders
      freshness_max: 60s
    - source: kyc.supplier_registry
      freshness_max: 24h
  memory: episodic
  freshness_default: 5m

action_space:
  tools:
    - name: erp.create_po
      signature: "(amount: EUR, supplier_id: str) -> po_id"
    - name: erp.send_to_approver
      signature: "(po_id: str, approver_group: str) -> ticket_id"
    - name: kyc.lookup_supplier
      signature: "(supplier_id: str) -> SupplierRecord"
  max_plan_length: 8
  cost_model:
    erp.create_po: { compute: low, external: medium }
    erp.send_to_approver: { compute: low, external: low }
    kyc.lookup_supplier: { compute: low, external: low }

reward:
  type: scalarization
  components:
    - name: throughput
      weight: 0.6
    - name: cycle_time
      weight: 0.4
  horizon: daily
  asymmetry:
    false_positive_cost: 1.0   # blocking a valid PO
    false_negative_cost: 50.0  # admitting an invalid PO
  goodhart_check:
    metric: throughput_vs_audit_pass_rate_correlation
    threshold: 0.7

constraints:
  - id: c14_human_oversight_high_value
    source:
      type: regulatory
      reference: "EU AI Act Art. 14"
    class: escalation
    predicate:
      lang: cel
      expr: "action.tool == 'erp.create_po' && action.args.amount >= 50000"
    operating_point:
      type: deterministic_threshold
      theta: 50000
      false_positive_tolerance: 0.01    # boundary cases at exactly EUR 50,000
      false_negative_tolerance: 0.00
    verification:
      point: PAG
      latency_budget_ms: 5
    response:
      type: suspend_and_route
      router_group: procurement_managers
    timeout:
      reversibility_window_s: 600
      on_timeout: deny    # default-deny; never default-allow under load
    trace_fields:
      - principal
      - planner
      - executor
      - tool
      - authority
      - constraint_id
      - operator_decision

  - id: ch_security_supplier_kyc
    source:
      type: regulatory
      reference: "EU AI Act Art. 15 (cybersecurity)"
    class: hard
    predicate:
      lang: cel
      expr: "supplier.kyc_status == 'verified' && !supplier.sanctions_match"
    operating_point:
      type: exact_predicate
      false_positive_tolerance: 0.00
      false_negative_tolerance: 0.00
    verification:
      point: tool_layer
      tool: erp.create_po
    response:
      type: block
    trace_fields:
      - principal
      - tool
      - constraint_id
      - kyc_status

  - id: cs_window_spend
    source:
      type: operational
      reference: "Finance Policy 4.2 (rolling spend cap)"
    class: soft
    predicate:
      lang: cel
      expr: "rolling_24h_spend(actor=principal) + action.args.amount <= 500000"
    operating_point:
      type: threshold
      theta: 475000                     # throttle threshold below the 500,000 cap
      calibration_basis: rolling_24h_spend
      false_positive_tolerance: 0.05
      false_negative_tolerance: 0.02
    verification:
      point: PAA
      latency_budget_ms: 10
    response:
      type: throttle
      backoff:
        formula: "exp(min(overage / 50000, 5))"
        unit: seconds
    trace_fields:
      - principal
      - rolling_24h_spend
      - overage_amount

  - id: ce_first_time_supplier
    source:
      type: contractual
      reference: "Master Procurement Agreement, Annex C"
    class: escalation
    predicate:
      lang: cel
      expr: "supplier.first_seen_at == null || age(supplier.first_seen_at) < 90d"
    operating_point:
      type: exact_predicate
      false_positive_tolerance: 0.00
      false_negative_tolerance: 0.00
    verification:
      point: PAG
      latency_budget_ms: 5
    response:
      type: suspend_and_route
      router_group: vendor_governance
    timeout:
      reversibility_window_s: 14400
      on_timeout: deny
    trace_fields:
      - principal
      - supplier_id
      - first_seen_at

escalation_router:
  groups:
    procurement_managers:
      capacity_model: { type: mmc, c: 2, mean_service_s: 360 }
      hours: "Mon-Fri 09:00-18:00 CET"
      after_hours:
        mode: emergency_on_call
        fallback_if_unavailable: deny     # honors c14 600s reversibility window
    vendor_governance:
      capacity_model: { type: mmc, c: 1, mean_service_s: 1800 }
      hours: "Mon-Fri 09:00-17:00 CET"
      after_hours:
        mode: defer_if_within_reversibility_window
        fallback_if_exceeds_window: deny  # honors ce 14400s reversibility window

audit_emission:
  schema_version: "sarc-trace-0.1"
  fields:
    state: [pre, post]
    action: [tool, args, plan_index]
    attribution: [principal, planner, executor, authority, capability]
    constraints_evaluated: [id, class, fired, response_taken]
    reward_components: [name, value]
  retention: "deployment-specific; set per applicable legal and internal retention policy"
  destination: append-only-store://compliance/procurement-approver

enforcement_property:
  property: "no_bypass"
  description: "All executable actions traverse at least one enforcement point compatible with their constraint class"
  invariant_ref: "I7"   # see Section 3.5
\end{verbatim}
\end{small}

\paragraph{Reading the specification.} The YAML names every object the framework requires. Each constraint declares its regulatory source, its class, the predicate (in a constraint expression language; CEL is used illustratively), the verification point, the response, and the trace fields. The escalation router's queueing model is declared explicitly as M/M/c, making the operator-staffing assumption part of the specification rather than an unstated dependency. The audit emission schema is generated from the specification, so an auditor reviewing this YAML can determine, ex ante, what the runtime trace will contain.

\paragraph{Compilation target.} A SARC specification of this form is consumable by a runtime that implements the four enforcement points; the constraint-expression language (CEL or equivalent) is evaluable by standard policy-as-code engines. The framework does not mandate a particular language; it requires that the predicate be deterministically evaluable within the latency budget declared at each enforcement point.

\section{Worked Economic Example: Operating-Point Asymmetry}\label{app:economic}

The economic argument of \S\ref{sec:economic} argues that constraint calibration is a capital-allocation decision under risk. We give here a small numerical illustration of how cost asymmetry drives the operating point toward conservative enforcement.

Suppose a deployment estimates the three cost coefficients of \S\ref{sec:economic} at $\kappa_{\text{FP}} = 1$ (a valid-action block costs one unit of throughput), $\kappa_{\text{ER}} = 5$ (an escalation costs five units of operator time), and $\kappa_{\text{FN}} = 100$ (a missed regulatory violation costs one hundred units of regulatory and reputational exposure). Under these coefficients, an operating point that increases the false-positive rate from 1\% to 5\% (a 4-percentage-point increase, costing $4 \kappa_{\text{FP}} = 4$ units per hundred actions) is economically justified if it reduces the false-negative rate by more than 0.04 percentage points (saving $> 0.04 \kappa_{\text{FN}} = 4$ units per hundred actions), before accounting for the corresponding change in escalation cost.

The point is not the specific numbers, which are illustrative; the point is the asymmetry. When $\kappa_{\text{FN}} \gg \kappa_{\text{FP}}$, even modest improvements in false-negative rate justify substantial increases in false-positive rate. Conservative enforcement is not over-cautious engineering in this regime; it is the economically rational allocation of risk under the declared cost structure. SARC's contribution is to require this declaration to be explicit (operating point $\theta$ as a field in each constraint, per invariant I2) rather than left as an unstated default chosen by whoever wrote the predicate.

\end{document}